\documentclass[twocolumn,showpacs,pre,floatfix]{revtex4-1}
\usepackage{ulem}% barrer souligner hachurer
\usepackage{verbatim}
\usepackage{graphicx}% Include figure files
\usepackage{epstopdf}% Include figure files
\usepackage{dcolumn}% Align table columns on decimal point
\usepackage{bm}% bold math
\usepackage{amsmath}
\usepackage{hyperref}
\usepackage{amssymb}
\usepackage[utf8]{inputenc}
\usepackage[T1]{fontenc}
\usepackage{mathtools}
\usepackage{upgreek}

\usepackage{xcolor}

\usepackage{enumitem}

\newcommand{\poro}{\Phi}
\newcommand{\deff}{\rho_{\text{eff}}}
\newcommand{\dwa}{\rho_{\text{w}}}
\newcommand{\dso}{\rho_{\text{s}}}
\newcommand{\din}{\rho_{\text{B}}}
\newcommand{\massin}{M_{\text{B}}}
\newcommand{\vin}{V_{\text{B}}}
\newcommand{\vinim}{V_{\text{B.im}}}
\newcommand{\vinem}{V_{\text{B.em}}}
\newcommand{\vnorm}{\mathcal{X}_{\text{in}}}
\newcommand{\rin}{r_{B}}
\newcommand{\h}{u}
\newcommand{\heq}{u_{\text{eq}}}
\newcommand{\hini}{u_{0}}

\newcommand{\gammagm}{\Gamma_{\text{H.L.}}}
\newcommand{\gammab}{\Gamma_{\text{G.E.L.}}}
\begin{document}

\setlength{\baselineskip}{0.4cm} \addtolength{\topmargin}{1.5cm}
\title{Sinking during earthquakes: critical acceleration criteria control drained soil liquefaction}
%\title{The art of sinking during earthquakes: critical acceleration criteria ruling drained soil liquefaction}
%\title{The art of sinking during earthquakes: the role of buoyancy in drained soil liquefaction}
%\title{Partial fluidization of saturated granular media through horizontal shaking}

\author{C. Cl\'{e}ment$^{1}$}
\author{R. Toussaint$^{1,2}$}
\email{renaud.toussaint@unistra.fr}
\author{M. Stojanova$^{3,1}$}
\author{E. Aharonov$^{4}$}

\affiliation{
$^1$Institut de physique du globe de Strasbourg,
Universit\'{e} de Strasbourg,
CNRS, UMR 7516, 67084 Strasbourg Cedex,
%Ecole et Observatoire des sciences de la Terre, 
%5 rue Rene Descartes, 67000 Strasbourg,
France.\\
$^2$PoreLab, Dept of Physics,
University of Oslo,
Norway.\\
$^3$
Institut Lumi\`{e}re Mati\`{e}re,
Universit\'{e} Lyon 1,
CNRS, UMR 5586, 69361 Lyon Cedex 07,
France.\\
$^4$
Hebrew University of Jerusalem,
Israel}

\date{\today}

\pacs{62.20.Mk, 46.50.+a, 81.40.Np, 68.35.Ct}

\maketitle

%%%%%%%%%%%%%%%%%%%%%%%%%%%%%%%%%%%%%%%%%%%%%%%%%%%%%%%%%%%%%%%%%%%%%%%%%%%%%%%%%%%%%%%%%%%%
%%%%%%%%%%%%%%%%%%%%%%%%%%%%%%%%%%%%%%%%%%%%%%%%%%%%%%%%%%%%%%%%%%%%%%%%%%%%%%%%%%%%%%%%%%%%
%%%%%%%%%%%%%%%%%%%%%%%%%%%%%%%%%%%%%%%%%%%%%%%%%%%%%%%%%%%%%%%%%%%%%%%%%%%%%%%%%%%%%%%%%%%%
%%%%%%%%%%%%%%%%%%%%%%%%%%%%%%%%%%%%%%%%%%%%%%%%%%%%%%%%%%%%%%%%%%%%%%%%%%%%%%%%%%%%%%%%%%%%
%%%%%%%%%%%%%%%%		ABSTRACT		%%%%%%%%%%%%%%%%%%%%%%%%%%%%%%%%%%%%
\section*{Abstract}
This article focuses on liquefaction of saturated granular soils, triggered by earthquakes.
%%% einat added -define liquefaction 
Liquefaction is defined here as the transition from a rigid state, in which the granular soil layer supports structures placed on its surface, to a fluid-like state, in which structures placed initially on the surface sink to their isostatic depth within the granular layer.
%%%
 We suggest a simple theoretical model for soil liquefaction and show that
buoyancy caused by the presence of water inside a granular medium has a dramatic influence on the stability of an intruder resting at the surface of the medium.
We confirm this hypothesis by comparison with laboratory experiments and Discrete Elements numerical simulations.
The external excitation representing ground motion during earthquakes 
is simulated via  horizontal sinusoidal oscillations of controlled frequency and amplitude.
In the experiments, we use particles only slightly denser than water,
 which as predicted theoretically, increases the effect of liquefaction and allows clear depth-of-sinking measurements.
In the simulations, a micromechanical model simulates grains using molecular dynamics with friction between neighbours.
The effect of the fluid is captured by taking into account buoyancy effects on the grains when they are immersed. 
We show that the motion of an intruder inside a granular medium
is mainly dependent on the peak acceleration of the ground motion, and establish a phase diagram for the conditions under which liquefaction happens, depending on the soil bulk density, friction properties,
presence of water, and on the peak acceleration of the imposed large-scale soil vibrations.
We establish that in liquefaction conditions,
most cases relax towards an equilibrium position following an exponential in time. 
We also show that the equilibrium position itself, for most liquefaction regimes,
corresponds to the isostatic equilibrium of the intruder
inside a medium of effective density.
The characteristic time to relaxation is shown to be essentially a function of the peak ground velocity.

%%%%%%%%%%%%%%%%%%%%%%%%%%%%%%%%%%%%%%%%%%%%%%%%%%%%%%%%%%%%%%%%%%%%%%%%%%%%%%%%%%%%%%%%%%%%
%%%%%%%%%%%%%%%%%%%%%%%%%%%%%%%%%%%%%%%%%%%%%%%%%%%%%%%%%%%%%%%%%%%%%%%%%%%%%%%%%%%%%%%%%%%%
%%%%%%%%%%%%%%%%%%%%%%%%%%%%%%%%%%%%%%%%%%%%%%%%%%%%%%%%%%%%%%%%%%%%%%%%%%%%%%%%%%%%%%%%%%%%
%%%%%%%%%%%%%%%%%%%%%%%%%%%%%%%%%%%%%%%%%%%%%%%%%%%%%%%%%%%%%%%%%%%%%%%%%%%%%%%%%%%%%%%%%%%%
%%%%%%%%%%%%%%%%		INTRODUCTION		%%%%%%%%%%%%%%%%%%%%%%%%%%%%%%%%%%%%
\section*{Introduction}

%Luding1996:
%Cite des références de review dans l'intro [1] et [47] 
%[1] milieux granulaire et ségrégation
%[47] molecular dynamics, les forces possibles
%grains non sphériques en référence [39].

Under usual conditions, natural and artificial soils (used as geotechnical foundations or construction materials) support the weight of infrastructure placed on their surface, and the stresses exerted on their surface are transmitted to the underlying grains along force chains \cite{Cates1998}.
However contacts between grains may be weakened during shaking, and/or by addition of a liquid phase, which in general exerts an additional fluid pressure on the grains. When these contacts break or slide, the system is not stable anymore,
so that the granular medium loses its ability to support shear stress and flows as a liquid, which is referred to as liquefaction \cite{Wang2010}. 
In such cases debris flows, avalanches, quicksands or liquefaction can occur. Buildings on liquefied soils may sink or tilt, and pipelines are displaced or float to the surface. All of the above phenomena may  lead to significant damage.

In this paper we focus on soil liquefaction associated with earthquakes \cite{Wang2010,ambraseys1969liquefaction,Huang2013}.
Some areas are well known to be prone to soil liquefaction,
like the New Madrid Seismic Zone in the central United States or
Mexico city in Mexico \cite{Wang2010,Diaz-Rodriguez1992}.
The last main earthquakes which have been followed by severe liquefaction effects
- listed in \cite{Huang2013} -
are
the 1964  Alaska Earthquake, magnitude Mw 9.2 \cite{Waller1966},
the 1964 Niigata Earthquake, magnitude  Mw 7.5  \cite{Kuribayashi1975,Seed1967}, Japan,
and the 2011 Christchurch Earthquake, magnitude Mw 6.3 \cite{Cubrinovski2011}, New-Zealand.

%%%%%%%%%%%%%%%%%%%%%%%%%%%%%%%%%%%%%%%%%%%%%%%%%%%%%%%% 
%% historical liquefaction
Liquefaction was historically first explained by Terzaghi \cite{Terzaghi1943}, relating liquefaction occurrence to the effective stress in the material.
Further geotechnical work \cite{Youd2001,Seed2003} improved the principe of Terzaghi 
in order to explain as many liquefaction cases as possible.
The current understanding of liquefaction, which underlies the construction principles for foundations and roads, can be summed up as follows:
%(une reference serait(terzaghi 1925 citée dans le bouquin \cite{Wang2010})).
During earthquakes, seismic waves disturb the grain-grain contacts,
and some weight initially carried by the sediments
are then shifted to the interstitial pore water \cite{Wang2010}.
The consolidation of the saturated sediment occurs in effectively undrained conditions (due to the short timescale of earthquakes),
and pore pressure builds up as the granular pack compacts.
As a result, the effective stress carried by the sediments decreases.
If the pore pressure rises more,
the solid weight can be entirely borne by pore water and the sediments become fluid-like, i.e. they cannot sustain shear stress in a static configuration.
% einat - i think we should say about thershold of liquefaction: complete loss of strength (what engineers think) vs sliding (what physics community thinks) 
This accepted mechanism thus assumes that the granular media  must lose its strength completely to produce liquefied behavior. 

%% i changed this paragraph - einat
The  above-described pore-pressure theory of earthquake-induced soil liquefaction and its recent advances indeed explain many natural instances of observed liquefaction \cite{Seed2003,Sawicki2006,Sawicki2009},
yet it fails to explain many other field observations of earthquake-induced liquefaction. Examples of types of field occurrences of liquefaction that are not explained by the pore-pressure theory \cite{Clement2016} include far-field liquefaction triggered at low energy density  \cite{Green2004,wang2007}, liquefaction under fully drained conditions \cite{Goren2010,Goren2011,Lakeland2014}, repeated liquefaction \cite{Obermeier1996} and liquefaction in pre-compacted soils \cite{Soga1998}.

%%%%%%%%%%%%%%%%%%%%%%%%%%%%%%%%%%%%%%%%%%%%%%%%%%%%%%%% 
%% paragraph on physics and grains 

%%added a clarifying sentence -einat
In fact, elevated pore-pressure is not the only path for granular material liquefaction. 
The phenomenon of solid-liquid transition of granular materials is known in a more general framework as fluidisation.
Fluidisation of a granular medium occurs when an initially rigid medium looses its cohesion and starts behaving like a fluid.
One of its most famous examples is quicksand -
a granular medium which can support a body on its surface until the said body is not moving,
whereas if it is moving, the body sinks into the quicksand  \cite{Khaldoun2005, Khaldoun2006}.

A compact, dense granular medium at rest behaves like a solid.
The grains experience friction due to the normal stress they apply to each other, traditionally coming from the gravitational loading.
The friction enables the grains to resist external forces without flowing \cite{Jiang2009}
and sustain weight by redistributing it along force chains \cite{Cates1998}.
The importance of normal stresses for the rigidity of a granular medium can be illustrated with some recent penetration experiments.
Lohse et al. in \cite{Lohse2004} and Brizinski et al. in \cite{BrzinskiIII2013a} used a homogeneous air injection
to loosen the granular medium and unload some of the gravitational force on grains.
After the injection was stopped, the granular medium could not carry the weight it was carrying before,
and any objects placed on its surface sank.
In a more extreme situation, where the grains are loose and also have a very low density (expanded polystyrene),
the penetration of the intruder is infinite, just like the penetration of a dense object in a liquid \cite{PachecoVazquez2011}.
The external energy required for fluidisation can be of different nature.
It can come, for example, from gravity forces \cite{Pouliquen1999, Bertho2003, Parez2016},
shear stress \cite{Losert2000}, vibrations \cite{Mujica1998, Huerta2005, Poeschel1995, Rosato1987, Khaldoun2006}
or the flow of an interstitial fluid \cite{Varas2013, Tsimring1999},
and produce granular flows that can exhibit fluid-like behavior such as buoyancy \cite{Huerta2005} or anti-buoyancy \cite{Shinbrot1998}
force, waves on their surface \cite{Mujica1998,PicaCiamarra2004}, flow instabilities \cite{Vinningland2007,Vinningland2007b,Vinningland2010, Vinningland2012EPJST, Niebling2010a,Niebling2010b}
and size segregation \cite{Poeschel1995,Rosato1987}.

%I added a clarifying sentence , re sliding as the criteria for liquefaction - einat. 
Granular media start to behave as fluids when global contact sliding initiates throughout the media. The force needed to initiate sliding depends on the strength of the grain contacts - for an easy fluidisation, one needs to unload some of the normal stress, hence reducing the friction forces.
For example, the shear stress necessary to make a granular layer flow decreases when the granular medium is vibrated \cite{Melhus2012},
the vibrations weakening the granular contacts. The presence of an interstitial fluid can play a similar role.
When the grains are immersed, the effective normal stress they are subjected to is lowered by the pore pressure of the fluid, hence reducing the friction forces between grains.
If the pore pressure is high enough, the friction forces between grains can be completely suppressed and the medium can not resist shear anymore \cite{Goren2011}.
Considerations on granular media have been used to generalise these liquefaction conditions on the heterogeneous pore pressure distribution in disordered granular media \cite{Goren2013}.
In another study, Geromichalos et al. \cite{Geromichalos2003} show that the addition of water (more than 1\% of the total volume)
decreases the segregation effect inside granular media subjected to horizontal shaking,
and attribute this to the fact that water makes the particles slide easier on each other.
%% I added the sentence below to summerize that sliding is the onset of liquefaction -einat
It is important to note that it is not necessary to unload all of the normal stress on contacts (with the prime example being pore pressure reaching the normal stress value) in order to initiate liquefaction. Instead, causing sliding of contacts throughout the media is sufficient to produce features of liquefaction. 

%%%%%%%%%%%%%%%%%%%%%%%%%%%%%%%%%%%%%%%%%%%%%%%%%%%%%%%%
%% description of our study

In this present study we will consider the effects of both water presence and vibrations, on the behavior of a granular medium.
The situation is very similar to the one in \cite{Sanchez-Colina2014}, where a dry granular medium was fluidised by vibration, and the sinking of an intruder initially placed on the granular surface was observed.
By shaking the granular medium to reproduce earthquakes
we can observe that objects originally resting on the surface partly or entirely sink in this medium.
The grain-grain contacts are disturbed by the shaking,
which allows some grains to slide on each other.
This effect is shown to be promoted by the presence of water, but does not require elevated pore pressure beyond the hydrostatic value.  
Our aim is to first highlight how liquefaction in such drained conditions can 
be explained by friction and sliding inside the medium,
and to characterise the liquefaction state according to the parameters of the shaking.
The first section presents the research questions,
the experimental material and a simple theoretical model for the phenomenon.
Section II  presents the methods for the experiments and simulations,
and the detailed characterisation of the liquefaction regimes.
Section III  presents the different results,
about the classification of deformation regimes as function of the applied shaking,
and about the characteristic sinking velocity and equilibrium depth.
Discussion of results and their consequences is presented in section IV.

%%%%%%%%%%%%%%%%%%%%%%%%%%%%%%%%%%%%%%%%%%%%%%%%%%%%%%%%%%%%%%%%%%%%%%%%%%%%%%%%%%%%%%%%%%%%
%%%%%%%%%%%%%%%%%%%%%%%%%%%%%%%%%%%%%%%%%%%%%%%%%%%%%%%%%%%%%%%%%%%%%%%%%%%%%%%%%%%%%%%%%%%%
%%%%%%%%%%%%%%%%%%%%%%%%%%%%%%%%%%%%%%%%%%%%%%%%%%%%%%%%%%%%%%%%%%%%%%%%%%%%%%%%%%%%%%%%%%%%
%%%%%%%%%%%%%%%%%%%%%%%%%%%%%%%%%%%%%%%%%%%%%%%%%%%%%%%%%%%%%%%%%%%%%%%%%%%%%%%%%%%%%%%%%%%%
%%%%%%%%%%%%%%%		PHYSICS OF LIQUEFACTION 		%%%%%%%%%%%%%%%%%%%%%%%%%%%%%%%%%%%%
\section{The physics of liquefaction}

The following section will provide an overview of the problem. 
We will qualitatively describe some of the experimental results in order to illustrate
the different behavioral regimes that we observed. 
We will then explain the mechanisms behind these behaviors and the transition between them,
and identify the link with soil liquefaction. 

\subsection{Description of the observed deformation regimes}

\begin{figure}[htbp]
\begin{center}
	\includegraphics[width=8cm]{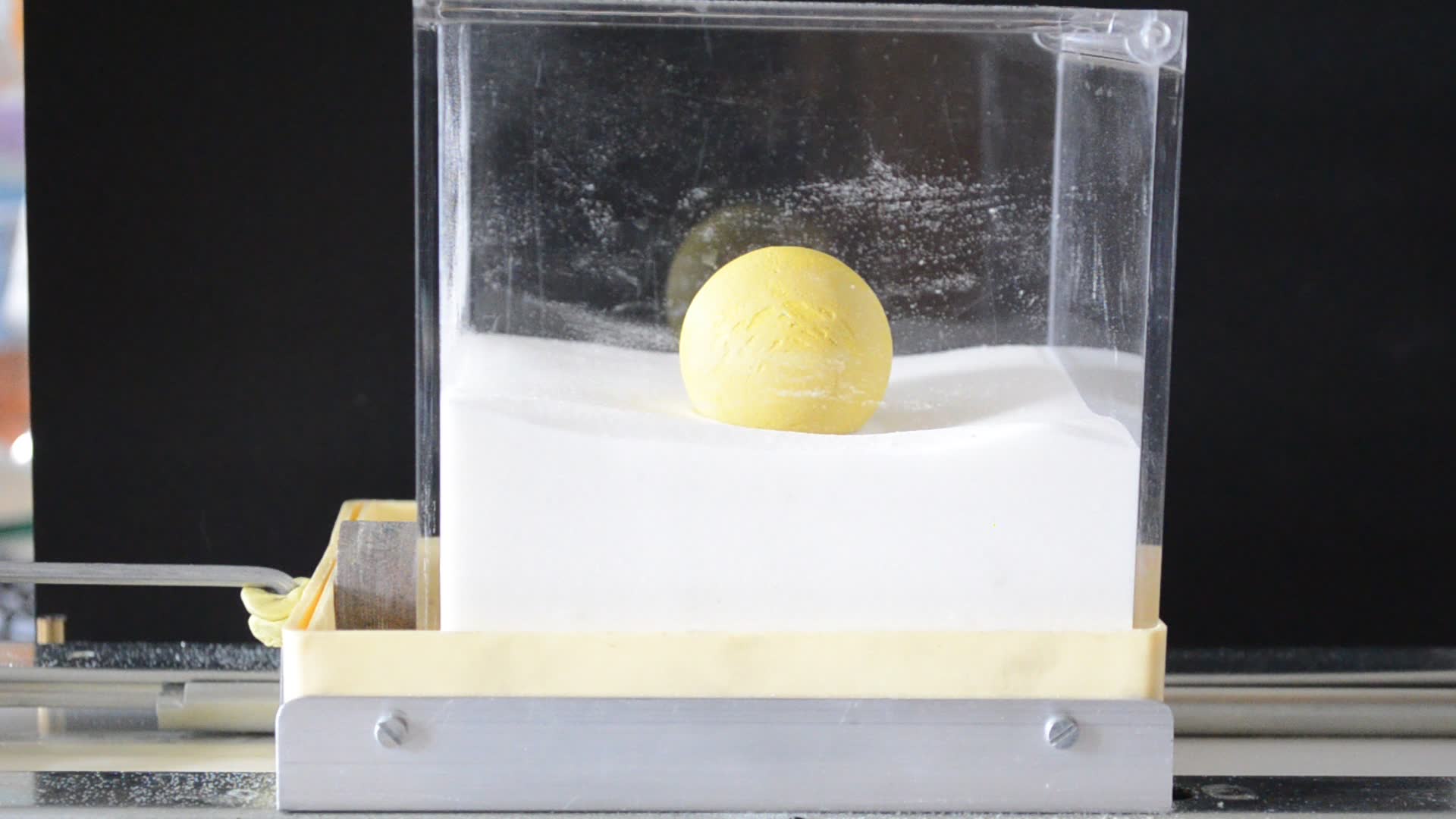}
	\caption[]{Initial state at mechanical equilibrium.
	The intruder diameter is 40 millimeters, its density is $1030 \pm 5 \mathrm{\ kg\,m^{-3}} $,
	close to the one of the grains composing the granular medium.
	The medium is either dry or fully saturated with water.
	This example shows a dry medium.}
	\label{im:ini_expe}
\end{center}
\end{figure}

Our experiment is a simplified model of a building resting on a soil during the passage of a seismic wave. 
The soil is simplified to a granular medium made of nonexpanded polystyrene spheres of density
$\rho_s=1050 \mathrm{\ kg\,m^{-3}}$ \cite{Microbeads} and mean diameter of $140 \mathrm{\ \upmu m}$.
It can be completely dry or completely saturated.
The granular medium is in a test cell, a transparent PMMA box of dimensions $12.8\mathrm{\ cm} \times 12.8cm\mathrm{\ cm} \times 12.5\mathrm{\ cm}$.  
A hollow sphere of 40 mm diameter,
and of effective density of $1030 \pm 5 \mathrm{\ kg\,m^{-3}}$
initially rests on the top of the layer (Fig.~\ref{im:ini_expe}), representing an analogue building.
To reproduce the effect of an earthquake we shake the different media horizontally with a controlled frequency and amplitude.
The frequency ranges from 0.15 Hz to 50 Hz, and the peak ground acceleration (PGA) from
$10^{-2} \mathrm{\ m\,s^{-2}}$ to $100 \mathrm{\ m\,s}^{-2}$,
corresponding to conditions met during earthquakes with macroseismic intensity of II to V-VI \cite{Souriau2006}.

\begin{figure}[htbp]
\begin{center}
	\includegraphics[width=8cm]{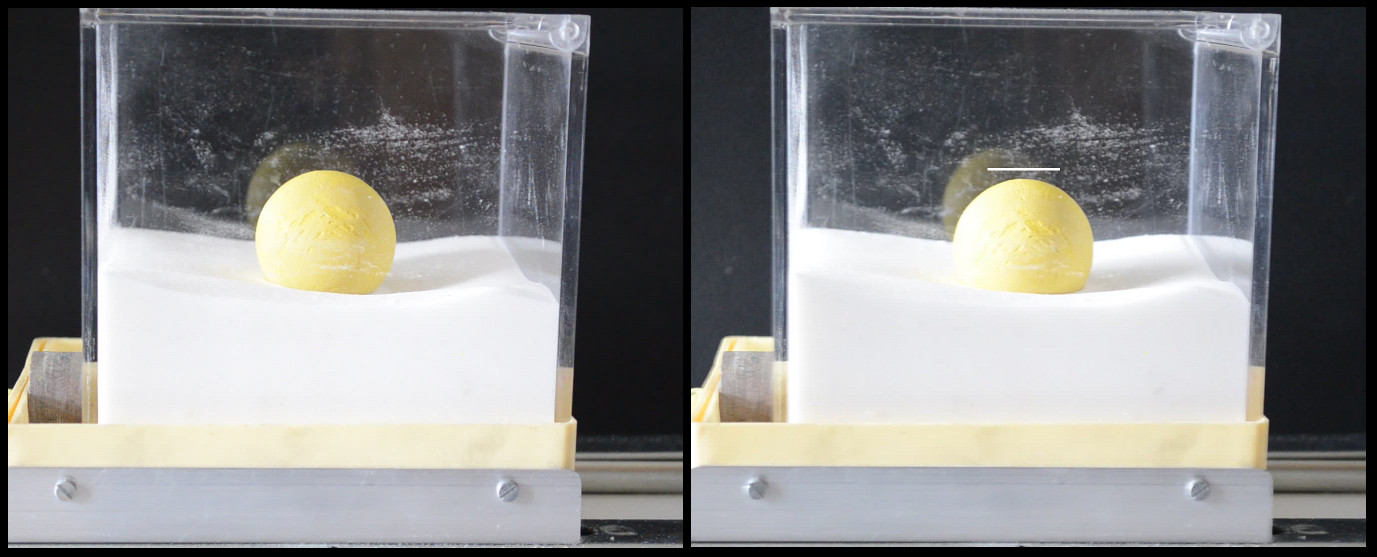}
	\caption[]{Dry medium, initial state on the left and final
          state on the right.
	The intruder sinks a few percent of its diameter.
	The initial position is represented on the final picture by a white horizontal line.
	The bulk densities of the intruder and the grains composing the medium are equal.}
	\label{im:sec}
\end{center}
\end{figure}

We observed that the behavior of the system depends on the PGA applied to it.
In the dry case, at small imposed accelerations the intruder and the particles follow the cell movement,
but are almost immobile with respect to each-other.
For larger PGA, convection cells appear inside the granular medium:
the particles on the top of the medium can be seen moving toward the sides.
The intruder stays at the surface, see Fig.~\ref{im:sec},
and can eventually roll from side to side if the acceleration is large enough. 

\begin{figure}[htbp]
\begin{center}
	\includegraphics[width=8cm]{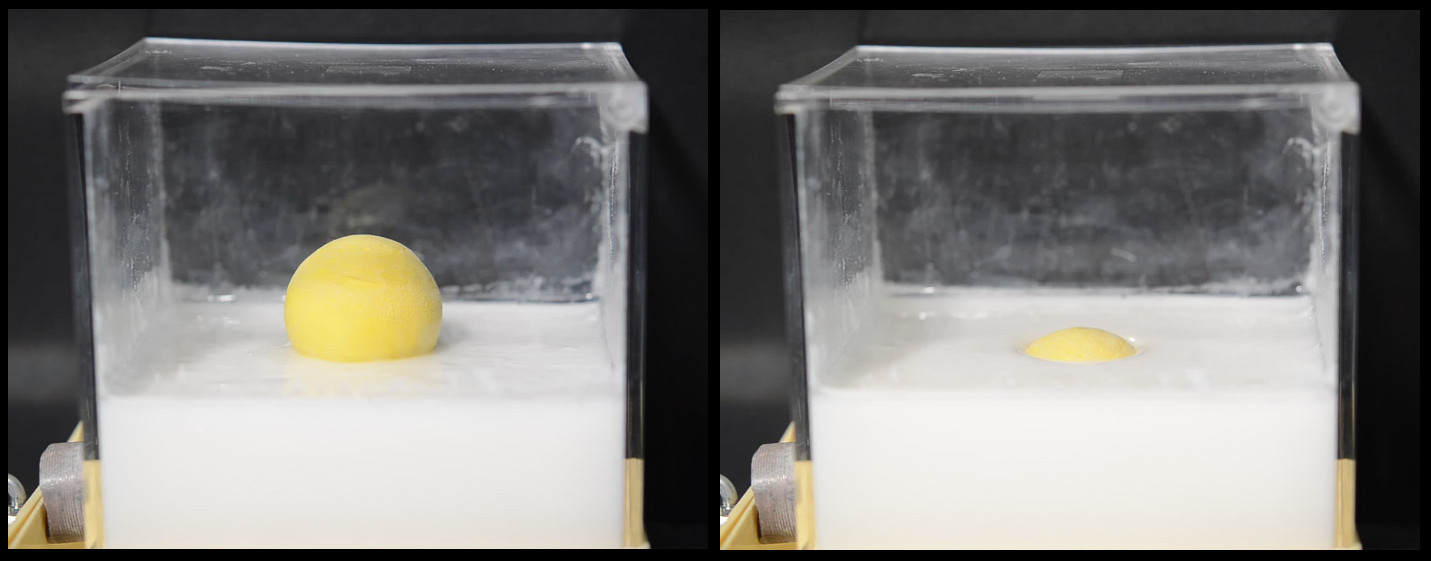}
	\caption[]{Saturated medium, initial state on the left and
          final state on the right.
	The intruder is eventually almost entirely immersed.
	The bulk densities of the intruder and of the grains composing the medium are equal.}
	\label{im:sature}
\end{center}
\end{figure}

With an initially saturated medium, still no significant motion is observed at low shaking accelerations.
However, when the acceleration is increased,
the intruder sinks rapidly into the medium, until an equilibrium is reached, 
which can be close to a total immersion, as shown on Fig.~\ref{im:sature}.
At the equilibrium the intruder remains almost immobile, 
and the rearrangements of the particles on the surface of the medium are too small to be observed.
For even larger imposed accelerations,
a similar  sinking of the intruder is observed, but accompanied by motion of the surrounding grains.
In this case the motion of the intruder and medium never ceases totally during the imposed oscillations.

For the saturated media, we can therefore identify three behaviors, occurring at different peak ground accelerations.
\begin{itemize}[leftmargin=*]
\item[]Low PGA : \textbf{Rigid} behavior \newline
If the acceleration of the medium is low, 
%i.e. of the order of few hundredths of the the Earth's gravitational acceleration,
%MENKA *** the quantitative study comes later. Is it necessary to have this here?***
%CECILE *** you are right, and there is a rough estimate of the acceleration needed to liquefy in the legend of figure 1  ***
the system oscillates like a solid, following the cell's movement.
The intruder stays at the surface and only a small descent of a few millimeters can sometimes be observed. 
\item[]Intermediate PGA : \textbf{Heterogeneous Liquefaction} behavior (H.L.)\newline
When the acceleration is increased beyond some critical level,
the intruder rapidly sinks in the saturated medium until attaining an equilibrium position where it stops moving.
The medium on the surface shows only little rearrangements.
As will be seen later in the simulations, the grains in the intruder
vicinity, underneath it, are in this case locked (not sliding on each
other), the contact normal stresses rising due to the intruder
weight. They accompany the motion of the intruder. In contrast, further from the intruder, the grains sometimes slide on each other - hence the term heterogeneous liquefaction, to reflect the difference between these two zones, i.e. the fact that the liquid-like behavior is heterogeneously distributed.
This behavior is only observed for a saturated granular medium.
In the dry case, at equal PGA the intruder stays on the surface of the granular medium. 
\item[]High PGA : \textbf{Global Excitation Liquefaction} behavior (G.E.L.)\newline
For even higher accelerations, we can observe a total and continuous rearrangement of the medium, presenting convection cells.
The intruder stays at the surface of the medium for dry cases,
or sinks in saturated conditions.
We call this behavior Global Excitation Liquefaction
because (as will be illustrated in the simulations) the whole medium
rearranges, sliding between grains can happen in the whole cell, and deformation never stops.
\end{itemize}

In our experiments, the solid behavior at low PGA corresponds to "regular" solid soil, sustaining the weight above it.
The G.E.L. behavior at high acceleration is not a phenomenon that is observed during earthquakes in Nature,
because it requires a very high acceleration,
and can only be reached during artificial excitation of granular material. 
In this case the fact that the intruder stays at the surface of dry granular media
is related to the Brazil nuts effect \cite{Luding1996,Clement2010,Poeschel1995,Geromichalos2003,Shinbrot1998}.
Finally, the H.L. behavior observed at intermediate PGA corresponds to soil liquefaction during an earthquake. 
In these experiments, it is the addition of water that enables the medium to liquefy.
Indeed, it is only when the medium is saturated that we observe a regime where an intruder can penetrate into it. 
The shape of the intruder also affects this behavior in dry grains,
since with similar densities, cylindrical objects can sink or tilt in
dry granular media \cite{Sanchez-Colina2016}.
%CECILE *** We want to keep this phrase to say that, even if we do not
%observe liquefaction into dry media, some colleagues of us observed liquefaction
%using a cylinder and the same granular material described here.

%The particles forming the experimental medium in our exemple are monodisperse and 
%15000 times smaller than the intruder.

%%%%%%%%%%%%%%%%%%%%%%%%%%%%%%%%%%%%%%%%%%%%%%%%%%%%%%%%%%%%%%%%%%%%%%%%%%%%%%%%%%%%%%%%%%%%
%%%%%%%%%%%%%%%%%%%%%%%%%%%%%%%%%%%%%%%%%%%%%%%%%%%%%%%%%%%%%%%%%%%%%%%%%%%%%%%%%%%%%%%%%%%%
\subsection{Problem definition and a simple model}
\label{section:problematic}
\begin{figure}[htbp]
\begin{center}
   \includegraphics[width=5cm]{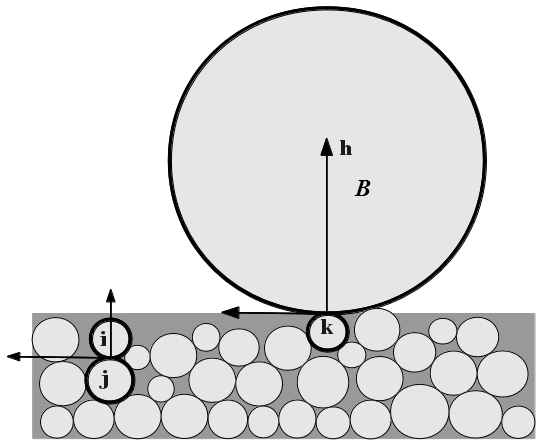}
   \caption
	[Theoretical model of saturated soil.]
	{Theoretical model of saturated soil with a spherical intruder on the top.
   The weight of particle $i$ applied on particle $j$ is reduced by the buoyancy
   whereas the weight of particle $B$ (the intruder), whose center sits at height $h$, is entirely transmitted to   particle $k$.}
   \label{im:theorytical_model}
\end{center}
\end{figure}

The observations described above highlight so far unreported aspects of liquefaction.
In our experiments, the presence of water is crucial for observing liquefaction-like sinking of the intruder.
We explain the physics of the liquefaction appearing in these experiments using
a simple theoretical soil consisting of a (saturated or dry) grain pack, as in Fig.~\ref{im:theorytical_model}.
This soil is composed of spherical particles and water filling the poral space between them.
A large sphere on  top of the granular soil represents a building built on it.
We assume that the situation is initially at mechanical equilibrium.
Here we will determine under which conditions this equilibrium can be broken,
and the large sphere could start to sink into the medium.
We first focus on the saturated cases, as represented on Fig.~\ref{im:theorytical_model}.

%%%%%%%%%%%%%	WITH WATER WATER WATER WATER  %%%%%%%%%%%%%
\subsubsection{With saturated medium}

%%%%%%%%%%%%%	Pi AND Pj %%%%%%%%%%%%%
To define the sliding condition we will distinguish the case of a contact between two grains
($P_i$ and $P_j$ on Fig. \ref{im:theorytical_model}), or between the intruder and a grain ($P_B$ and $P_k$). 
First consider two particles inside the saturated soil $P_i$ and $P_j$ placed on top of each other.
The normal force at the contact  $ij$ acting on the lower sphere $P_j$ is
the effective weight of the column above it resulting from gravity and buoyancy force.
\begin{eqnarray}
\label{eq:fnij}
\boldsymbol{F_n^{ij}}=M_{above} (1-\frac{\dwa}{\dso}) \boldsymbol{g},
\end{eqnarray}
 where $M_{above}$ is the mass of the grains inside the column above $P_j$,
$\dso$ the particle density, $\dwa$ the water density
and $ \boldsymbol{g}$ is the gravitational acceleration.
Next,  consider the intruder $P_{B}$- "B" for building-
and the particle of the soil right under it (or the set of grains under it and in contact with it) $P_k$.
The normal force that $P_{B}$ applies on  $P_k$ is
\begin{eqnarray}
\label{eq:fnbj}
 \boldsymbol{F_n^{Bk}}=\massin  \boldsymbol{g},
\end{eqnarray}
with $\massin$ the mass of $P_{B}$.
There is no buoyancy term in this case, since no part of the intruder
is submerged under water.
We apply to this soil a horizontal oscillation with a lateral displacement of the form 
$A\sin(\omega t)$.
The peak ground acceleration due to this movement is therefore $A\omega^2$.
We consider that the medium and intruder follow the imposed external motion,
in order to check whether the contacts reach a sliding threshold, and use this as a sign of possible deformation.
At small acceleration, the contact $ij$ is rigid and 
experiences a tangential force of the form 
\begin{eqnarray}
\label{eq:ftij}
 \boldsymbol{F_t^{ij}}=M_{above} A\omega^2.
\end{eqnarray}
We assume that each contact follows a Coulomb friction law,
where $\mu$ is the internal friction coefficient equal to the tangent of the repose angle
of the considered granular material.
Thus if the tangential force on the contact exceeds the criterion set by the Coulomb friction law,
the contact $ij$ slides and
Eq.~(\ref{eq:ftij}) becomes $ \boldsymbol{F_t^{ij}}=\mu \boldsymbol{F_n^{ij}}$.
Thus the medium remains rigid if 
\begin{eqnarray}
\nonumber
|M_{above} A\omega^2| < |\mu \boldsymbol{F_n^{ij}}|
\end{eqnarray}
i.e. while
\begin{eqnarray}
\nonumber
M_{above} A\omega^2 < \mu M_{above} (1-\frac{\dwa}{\dso})g.
\end{eqnarray}
Now, if we introduce a dimensionless peak ground acceleration,
normalized by the gravitational acceleration, $\Gamma=\displaystyle \frac{A\omega^2}{g}$,
the previous equation becomes:
\begin{eqnarray}
\label{eq:pipj}
\displaystyle \Gamma < \mu\frac{\dso-\dwa}{\dso}.
\end{eqnarray} 
While $\Gamma$ is low enough to satisfy Eq.~(\ref{eq:pipj}),  the particles inside the saturated soil
don't slide  on each other, and the medium acts like a rigid body.
\\

%%%%%%%%%%%%%	PB AND Pk %%%%%%%%%%%%%
The condition for sliding of the contact between the intruder $P_B$ and particle under it $P_k$ is different:
The horizontal oscillation induces a tangential force on $P_B$,
which will slide on $P_k$ if and only if 
\begin{eqnarray}
\nonumber
M_B A\omega^2 > |\mu \boldsymbol{F_n^{Bk}}|=M_B \mu g.
\end{eqnarray}
In other words, the emerged particle $P_B$ will stick on $P_k$ while
\begin{eqnarray}
\label{eq:pbpk}
\Gamma< \mu.
\end{eqnarray}
If $\Gamma> \mu$ the intruder can slide on the particle below it.
We can see in Eq.~(\ref{eq:pbpk}) that the acceleration $\Gamma$ required for the intruder to slide is higher
than the one needed to make the immersed particles of the soil slide, Eq.~(\ref{eq:pipj}).
This is due to the presence of water which carries a non negligible
part of the particles weight through the buoyancy,
so that the solid pressure between them is reduced and they can slide more easily.
The emerged intruder is not partially carried by water and its
contact on particle $P_k$ is stronger.
Depending on $\Gamma$ and according to the previous results,
three different regimes can be defined for the granular system:

\begin{eqnarray}
\label{eq:rig}
\text{\textbf{Low $\Gamma$}}&\text{\textbf{: }}&\displaystyle\Gamma< \mu\frac{\dso-\dwa}{\dso}\\
\label{eq:liq}
\text{\textbf{Intermediate $\Gamma$}}&\text{\textbf{: }} &\displaystyle\mu\frac{\dso-\dwa}{\dso}<\Gamma< \mu\\
\label{eq:flu}
\text{\textbf{High $\Gamma$}}&\text{\textbf{: }} & \displaystyle\mu<\Gamma
\end{eqnarray}

For low accelerations, the tangential force resulting on the particles contacts is too low to make any particle slide;
the medium can not rearrange, and will behave like a solid. 
Hence, this regime corresponds to the solid behavior of the system observed during our experiments. 
For intermediate $\Gamma$, many of the small particles can slide on each-other, while the intruder cannot slide on the particles beneath it. 
Hence, we can reasonably assume that the intruder will sink downwards because the medium is rearranging around it, 
and that this regime corresponds to the Heterogeneous Liquefaction case (H.L.).
Finally, for $\Gamma>\mu$, the intruder can also slide on the particles beneath it, hence it behaves like the other particles.
One can reasonably suppose that it will not continuously sink in the medium,
because of the Brazil nut effect \cite{Luding1996,Clement2010,Poeschel1995,Geromichalos2003,Shinbrot1998}. 
If $\Gamma$ gets even larger, until satisfying $\Gamma>1$, the medium gets decompacted. 
The acceleration is then large enough to defy gravity and the particles can make short jumps (short ballistic trajectories above the connected medium).
The case of $\Gamma>\mu$ corresponds to the Global Excitation Liquefaction case (G.E.L.).

Even though our model of the liquefaction system is very simple, it predicts three distinct regimes which may be identified experimentally. 
The rest of our work will focus on experimental and numerical verification of these predictions. 
We will systematically vary $\Gamma$ in order to explore the three regimes defined by this model.
The case we are mostly interested in for its representativity of natural liquefaction
during earthquakes is the case of Intermediate $\Gamma$,
where the submerged small particles can slide around a static intruder, making the intruder sink in the medium. 

In the following we will refer to the theoretical boundary between the three regimes by
\begin{eqnarray}
\gammagm=\mu\frac{\dso-\dwa}{\dso} \text{ and }
\gammab=\mu.
\label{eq:threshold}
\end{eqnarray}

%%%%%%%%%%%%%	WITHOUT WATER NO WATER NO WATER NO WATER  %%%%%%%%%%%%%
\subsubsection{With dry medium}
Inside dry granular media, the buoyancy forces disappear.
Since we initially considered an intruder emerged above the saturated granular medium,
Eq.~(\ref{eq:pbpk}) remains correct for dry media.
Eq.~(\ref{eq:pipj}), which gives the acceleration at which two particles of the soil can slide 
on each other, becomes 
\begin{eqnarray}
\label{eq:pipjdry}
\displaystyle \Gamma < \mu\frac{\dso-0}{\dso} = \mu,
\end{eqnarray}
which is identical to Eq.~(\ref{eq:pbpk}). 
Hence, in the case of a dry granular medium,
the sliding conditions are the same for the particles of the medium and for the intruder,
provided that the friction coefficient $\mu$ is the same for grain-grain contacts and for intruder-grain contacts.
The Intermediate acceleration case given by Eq.~(\ref{eq:liq}) disappears for dry media.
With an increase of $\Gamma$, this theory predicts that dry media will change their behaviors from the
Rigid case to the G.E.L. case around 
\begin{eqnarray}
\gammagm\,=\, \gammab\,=\,\mu.
\label{eq:thresholddry}
\end{eqnarray}

%%%%%%%	 	V isostatic		%%% %%%
\subsubsection{Final intruder position in a saturated medium}
\label{theoretical_final_position}
Let us consider next the final equilibrium state reached by the intruder in the saturated medium during liquefaction regimes.
Assuming that vertical friction forces average to zero, and only buoyancy forces and gravity dictate the final depth, 
the final position of the intruder can be estimated as the isostatic depth of the intruder inside a fluid of effective density $\deff$,
taking into account the particle density,
the fluid density and the porosity $\poro$.
We define the effective medium density $\deff$ as $\deff=\poro \dwa+(1-\poro) \dso$.
We measure $\poro$ in our experiments to be between $0.345 $ and $0.365 $,
which is close to a close random pack density \cite{Scott1969}, so that
%-- RENAUD: YOU SHOULD CHECK WITH YOUR MEASUREMENTS. IT IS CERTAINLY CLOSE, BUT I STILL SOMEWHERE BETWEEN LOOSe AND CLOSE. HOW MUCH DID YOU MEASURE, WITH WAt ERROR BAR? YOU SHOULD ALSO GIVE THE INITIAL POROSITY MEASURED IN THE NEXT SECTION -- 
%its porosity is around $0.365 \pm0.05$, REF NEEDED
%and FORMULA rhowater+(rhobuk-rhowater)(1-phi) SHOULD BE PRECISED FIRST
$\deff=  1032.2 \pm 0.4 \mathrm{\ kg\,m^{-3}}$.
If the final pressure profile in the granular medium is identical to a simple hydrostatic fluid situation, and if the medium acts as an effective viscous fluid,
the motion of the intruder is ruled by the following equation:
\begin{eqnarray}
\vin\din g   - V_{\text{B.im}}(z)\deff g  -\alpha(z)  \dot{z}= \vin\din g  \ddot{z} ,
\label{eq:intruderdyn}
\end{eqnarray}
where $ z$ is the downwards pointing vertical coordinate of the center of the intruder,
$V_{\text{B.im}}(z)$ is the immersed volume of the intruder (depending on its elevation),
$\din$ is the intruder density and $\vin$ its total volume.
The first term of Eq.~(\ref{eq:intruderdyn})  refers to the weight of the intruder, and the second
term refers to the buoyancy force.
Finally, $\alpha  \boldsymbol{\dot{z}}$ is a dissipative term due to forces exerted by the particles on the intruder, slowing down its motion. 
In the case of an effective medium of density $\deff < \din$ the intruder
is supposed to sink continuously because it is denser than the effective medium, while if $\deff > \din$ it reaches an equilibrium set by isostasy.
Since the intruder density is chosen as $1030 \mathrm{\ kg\,m^{-3}}$ in experiments,
% and $1000 \mathrm{\ kg\,m^{-3}}$ in simulations,
a macroscale equilibrium state exists with these simple assumptions,
and the intruder is expected to sink until it is nearly entirely immersed.
If this state is reached, we name $\vinim^{\text{equilibrium}}$
the immersed volume of the intruder under isostatic equilibrium, corresponding to:
\begin{eqnarray}
\nonumber
\vin\din g - \vinim^{\text{equilibrium}}\deff g =0,
\end{eqnarray}
giving
\begin{eqnarray}
\vinim^{\text{equilibrium}}=\vin\frac{\din}{\deff}.
\label{eq:vinimeq}
\end{eqnarray}
This value will be used as a theoretical reference and compared to
the final immersed volume observed in our experiments and simulations.
%If the solid pressure vanished definitely between the intruder and its surrounding particles,
%the intruder is entirely born by the fluid.
%%%%%%%%%%%%%%%%%%%%%%%%%%%%%%%%%%%%%%%%%%%%%%%%%%%%%%%%%%%%%%%%%%%%%%%%%%%%%%%%%%%%%%%%%%%%
%%%%%%%%%%%%%%%%%%%%%%%%%%%%%%%%%%%%%%%%%%%%%%%%%%%%%%%%%%%%%%%%%%%%%%%%%%%%%%%%%%%%%%%%%%%%
%%%%%%%%%%%%%%%%%%%%%%%%%%%%%%%%%%%%%%%%%%%%%%%%%%%%%%%%%%%%%%%%%%%%%%%%%%%%%%%%%%%%%%%%%%%%
%%%%%%%%%%%%%%%%%%%%%%%%%%%%%%%%%%%%%%%%%%%%%%%%%%%%%%%%%%%%%%%%%%%%%%%%%%%%%%%%%%%%%%%%%%%%
%%%%%%%%%%%%%%%%		METHODES 		%%%%%%%%%%%%%%%%%%%%%%%%%%%%%%%%%%%%
\section{Experimental and simulation methods for tracking liquefaction}
%%%%%%%%%%%%%%%%		EXPERIMENTS 		%%%%%%%%%%%%%%%%%%%%%%%%%%%%%%%%%%%%
\subsection{Presentation of the experiments}

Our experiments consist of following the movement of an intruder as it sinks into a
liquefied granular medium.
%The intruder used is a ping pong ball, partially filled with lead beads to control its density.
The intruder is a spherical ball, $4 \mathrm{\ cm}$ in diameter. 
We used the 123D\textregistered~ Design software in order to design the ball
and printed it with a MakerBot\textregistered~ Replicator2X 3D printer.
The sphere is made of heated polymeric material: an Acrylonitrile butadienestyrene (ABS) filament (type "color true yellow").
Designing our own balls,
we are able to control their effective density by adjusting the thickness of the shell layer, leaving a concentric empty sphere in the center --  without
adding any extra weight in the spherical shell, which allows to keep the spherical symmetry of the intruder density.
The granular medium is made of water and monodisperse spherical polystyrene beads,
with a diameter of $140 \mathrm{\ \upmu m}$ (DYNOSEEDS\textregistered~ TS \cite{Microbeads})
and density of $1050 \mathrm{\ kg\,m^{-3}}$.
The friction coefficient of this material $\mu_{\text{exp}}$  is estimated at  $\mu_{\text{exp}}=0.48$
by measuring the angle at which a thick homogeneous layer of the material starts to slide.

\begin{figure}[htbp]
  \begin{center}
  \includegraphics[width=7cm]{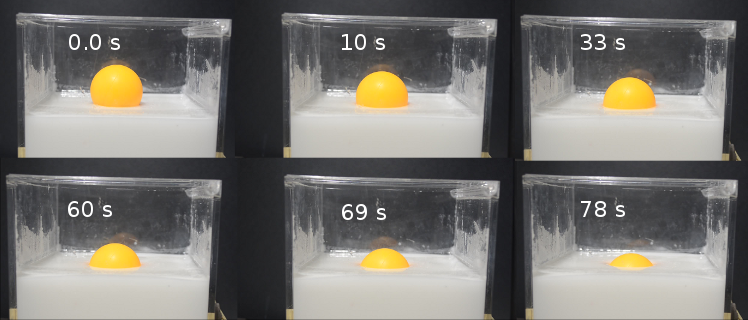}
  \caption
	[Series of snapshots of an experiment]
	{Series of snapshots of an experiment. Read from the left to the right and from the top to the bottom.}
  \label{fig:snap}
  \end{center}
\end{figure}

The experiments shown in this paper
used an intruder of density $1030 \pm 5 \mathrm{\ kg\,m^{-3}}$.
The experimental protocol is as  follows:
first we introduce water in a transparent PMMA cubical box of dimensions $12.8 \mathrm{\ cm}$ $\times$ $12.8 \mathrm{\ cm}$ $\times$ $12.8 \mathrm{\ cm}$.
We roughly fill the box up to a third of the desired final height.
We next let the polystyrene beads  rain down from random positions into the water, using a sieve, until the top of the beads piling up at the bottom of the container
reaches the surface of the water.
\begin{figure}[htbp]
\begin{center}
   \includegraphics[width=7cm]{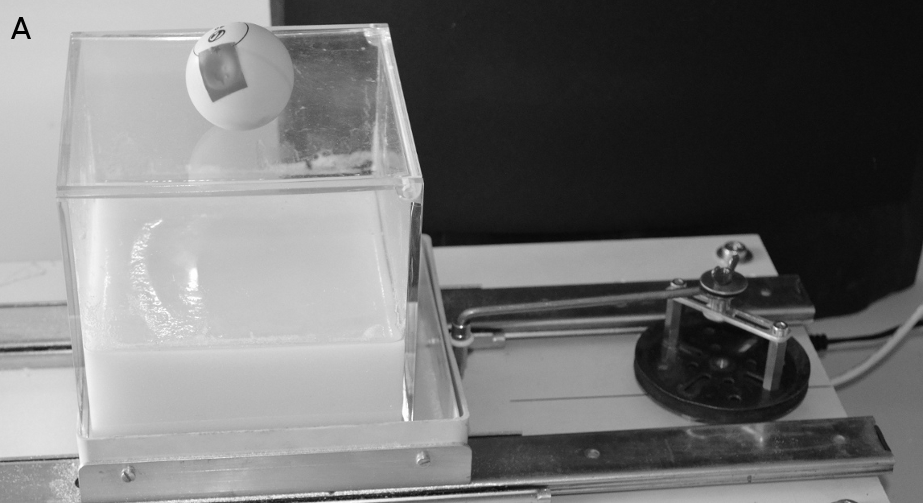}\\
   \includegraphics[width=7cm]{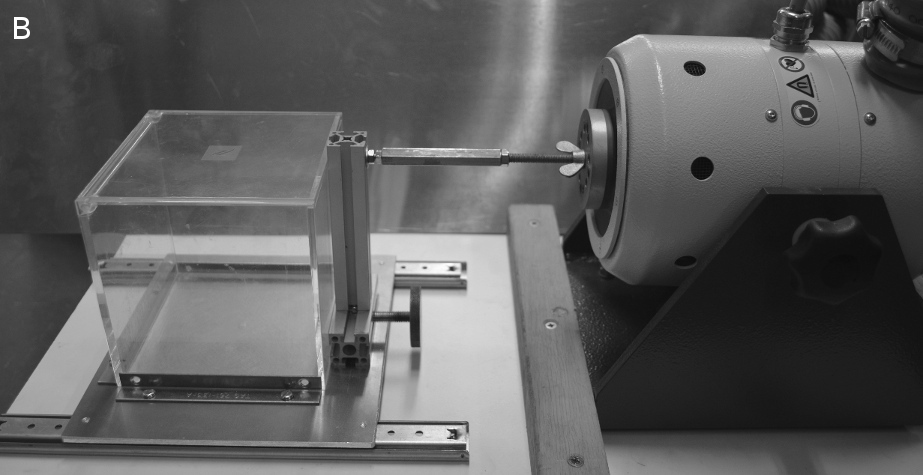}
\end{center}
   \caption[Experimental setups.]
   {Experimental setups. The mechanical part on the right exerts a horizontal movement on the box, guided by the rails.
A: Home-developed vibrator, using a Phidget\textregistered~1063 PhidgetStepper Bipolar 1
and Matlab\textregistered~ controls. 
This stepmotor provides an oscillation  with an amplitude range (mm) $[5; 30]$ and  a frequency range (Hz) $[0.15; 2.8]$.
B: TIRA\textregistered~ TV51120 shaker,
that we used with an amplitude range (mm) $[0.2; 1.5]$ and frequency range (Hz) $[4; 100]$.}
   \label{fig:setup}
\end{figure}
Two versions of the setup are shown in figure \ref{fig:setup}, using two different vibrators reaching different powers and frequency ranges: A. a home made vibrator, using a Phidget\textregistered~ 1063 PhidgetStepper Bipolar 1 and Matlab\textregistered~ controls, and B. a TIRA\textregistered~ TV51120 shaker, type
S51120, for higher frequencies and larger power.
After 3 minutes of relaxation time, sufficient for the granular matter to settle in the wet medium,
we gently depose the intruder on the surface of the medium.
After another minute of relaxation,
the box is horizontally shaken  with a sinusoidal movement
of controlled amplitude and frequency.
A camera  records the experiments.
In setup A of figure \ref{fig:setup} we use a Nikon\textregistered~ Digital Camera D5100 with a 80 mm objective
recording at 25 frames per second.
In  setup B we use a fast camera Photron\textregistered~ SA5 with a similar objective at 20000 frames per second.
The setup is illuminated by a flickerfree HMI 400 W Dedolight\textregistered~ spotlight in front of the experimental cell, next to the camera.
The videos are cut into series of snapshots using the free software FFmpeg\textregistered.
Figure \ref{fig:snap}  presents six snapshots,
corresponding to the different positions of the intruder
from the beginning to the end of the shaking.
We can follow the position of the intruder inside the medium through image analysis.
We use Matlab\textregistered~ algorithms and based on the color of each pixel of each picture,
we access the position of the pixel of the highest point of the ball.
Using these data and geometrical considerations to correct for perspective effects, we obtain the height 
of the ball above the granular medium surface.

%%%%%%%%%%%%%%%%%%%%%%%%%%%%%%%%%%%%%%%%%%%%%%%%%%%%%%%%%%%%%%%%%%%%%%%%%%%%%%%%%%%%%%%%%%%%
%%%%%%%%%%%%%%%%%%%%%%%%%%%%%%%%%%%%%%%%%%%%%%%%%%%%%%%%%%%%%%%%%%%%%%%%%%%%%%%%%%%%%%%%%%%%
%%%%%%%%%%%%%%%%		SIMULATIONS 		%%%%%%%%%%%%%%%%%%%%%%%%%%%%%%%%%%%%
\subsection{Numerical simulations}
\subsubsection{Modelling principles}
Our simulations are two dimensional (2D) representations of the experimental setup,
based on discrete element method (DEM) of molecular dynamics \cite{allen1989computer}.
We use the soft-particles approach originally developed by Cundall and Strack
\cite{cundall1979discrete} where we add a buoyancy force to account for the presence of water \cite{Niebling2010a}.
The simulations give access to the
trajectory and transient forces acting on individual 
cylindrical particles immersed in a fluid inside a finite space.
In order to model a 2D space of size comparable to the experiments,
we need to use larger grains than the experimental ones, since the experiments performed include roughly $10^8$ particles,
%diametre=0.140 mm
%volume = 4/3*3.14*0.140**3 = 0.011488
%volume boite = 120*120*100 = 1440000
%nb billes =  volume boite*0.64/volume = 80221374
which is beyond numerical capabilities of the model described here.
The behavior of each particle of mass $m$ and moment of inertia $I$ is governed by
the second law of Newton and the angular momentum theorem:
%einat - i added the word angular 
\begin{eqnarray}
  \label{eq:newton}
  \sum{ \boldsymbol{F}_{\text{ext}}}  &  =  & m \boldsymbol{\ddot{z}}(t) \\ 
  \sum{\mathcal{M}( \boldsymbol{F}_\text{{ext}})} &  =  &  \boldsymbol{I}\frac{\text{d}\dot{\theta}}{\text{dt}}(t), \nonumber
\end{eqnarray}
where $\sum{ \boldsymbol{F_{\text{ext}}}}$ and $\sum{\mathcal{M}(
  \boldsymbol{F}_\text{{ext}})}$ are the sum of external forces and
sum of external torques acting on the particle, repectively. 
% einat - i added the above senetnce  to define torques 
$ \boldsymbol{\ddot{z}}(t)$ is the particle acceleration
and $\dot{\theta}(t)$ is its angular velocity.
Our particles are cylinders because our simulation is in 2D,
thus for a particle of radius  $r$ 
the inertial momentum is  $I=mr^2/2$,
and the mass is $m=\dso \pi r^2 l$
where $l$ is the size of the medium in the third direction.
%We take arbitrarily $h=0.1 r_{\text{mean}}$.
To reproduce the experimental setup, the numerical media are enclosed between walls,
two vertical ones on each side and a horizontal one on the bottom (Fig.~\ref{fig:simu}).

%%%%%%%%%%%%%%% FORCES %%%%%%%%%%%%%%%%%%%%
We compute the forces in the Galilean laboratory reference frame.
The forces implemented on each particle are the gravity,
the buoyancy force of the liquid, and
the contact forces.
We assume the movement of the fluid with respect to the grains to be slow enough to neglect
the viscosity of the fluid.
Thus, the fluid only intervenes in this model  via buoyancy forces.
For a particle of density $\dso$, volume $V$ and immersed volume $V_{\text{im}}$,
the gravity and buoyancy forces are given respectively by 
$ \boldsymbol{F}_{\text{gravity}}=V\dso g  \boldsymbol{e}_z$ and
$ \boldsymbol{F}_{\text{buoyancy}}=- V_{\text{im}}\dwa g  \boldsymbol{e}_z $
where $g=9.81\mathrm{\ m\,s^{-2}}$ and
$ \boldsymbol{e}_z$ is the downwards vertical unit vector.
We model the contacts between two particles with a linear spring-dashpot model \cite{cundall1979discrete}.
For each contact we take into account a visco-elastic reaction
with two springs-dashpots, one in the normal direction and one in the tangential direction
in the local frame of the contact.
The springs exert a linear elastic repulsion,
with $k$ the elastic constant,
while the dashpot models exert a dissipative force during contact as a solid viscosity, 
i.e. viscous damping during the shocks,
with $\nu$ the viscosity.
The particles can rotate due to friction on contacts.
We implement a Coulomb friction law for each contact.
If the tangential force exceeds the Coulomb criterion,
we let the particle contact slide and set the tangential force equal to
the normal force times the friction coefficient.
%The laterals wall can move horizontally and the bottom can move vertically.
The interactions between  particles and  the three walls are the same as between
two particles, meaning that
the walls have similar mechanical and contact properties as the particles.
Once we have computed the sum of external forces for each particle,
we deduce their acceleration and use a leap-frog form of the Verlet algorithm \cite{allen1989computer} to get the velocity
and  position for the next timestep.
The particles positions and velocity are updated and we compute the new forces.

%%%%%%%%%%%%%%%%%%%%%%%%%%%%%%%%%% TIMESTEP %%%%%%%%%%%%%%%%%%%%

For a realistic model, the grains need to be hard and the overlaps small.  
According to the differential equations governing the system,
the duration of a contact is approximately given by $\sqrt{\frac{m}{k}}$,
so harder grains correspond to higher $k$ and to shorter collision durations.
The time step needs to be smaller than the collision duration, 
hence implementing harder grains means shorter time steps and longer computation times.
We need a time step at least 10 times smaller than this impact duration,
and we are interested in having the largest time step possible to reduce computational time, which means a small enough $k$.
%The timestep of the molecular dynamics simulations and the elasticity parameter have to be chosen carefully in this type of
%simulations, the former being dependent on the later.
%The timestep is chosen to be small enough with respect to all the other characteristic time, i.e. essentially the duration of the impact between two beads.
Simultaneously the elasticity parameter $k$ has to be large enough to avoid large deformations of the particles themselves.
Here we require these deformations not to exceed 1\%,
which physically translates in the contact force (solid stress times cross-section) being lower than $k\,0.01\,r_{\text{mean}}$.
The solid stress has a static and a dynamic component, the later appearing during impact only.
The static stress evolves in the medium as $\deff g z$ with $z$ the particle depth,
and the dynamic one evolves like $\dso v^2$ with $v$ the velocity of the particles. 
In our system the maximum value of solid stress is attained during high-velocity collisions,
when the static stress ($\deff g z$ with $z$ the particle depth) becomes negligible compared to the dynamic stress
($\dso v^2$ with $v$ the velocity of the particles). 
The maximal velocity of the particles is attained during the preparation stage and is around 1 m/s. 
Eventually we need to choose a value for $k$ such that $(\deff g z + \dso v^2) r_{\text{mean}} l < k\,0.01\,r_{\text{mean}}$
with $r_{\text{mean}}$ the mean radius of the set of particles.
We choose an elasticity coefficient of $k=20000 \mathrm{\ kg\,s^{-2}}$ 
and a timestep of $1. 10^{-6}\mathrm{\ s}$,
which suits all our simulations.
We checked that the value of the elasticity constant $k$ does not affect the behavior of
the media by doubling and quadrupling its value.

%%%%%%%%%%%%%%%		GRANULAR MEDIUM CREATION	%%%%%%%%%%%%%
\subsubsection{Our numerical granular media}
The first step is the creation of initial configurations.
We define $N_{MAX}$ the maximal number of particles of radius $r_{\text{mean}}$ which can fit
in the width of the box $w_{BOX}$:
\begin{equation}
N_{MAX} = \frac{w_{BOX}}{2\,r_{\text{mean}}}.
\label{N_max}
\end{equation}

\begin{figure}[htbp]
\begin{center}
	\includegraphics[width=6cm]{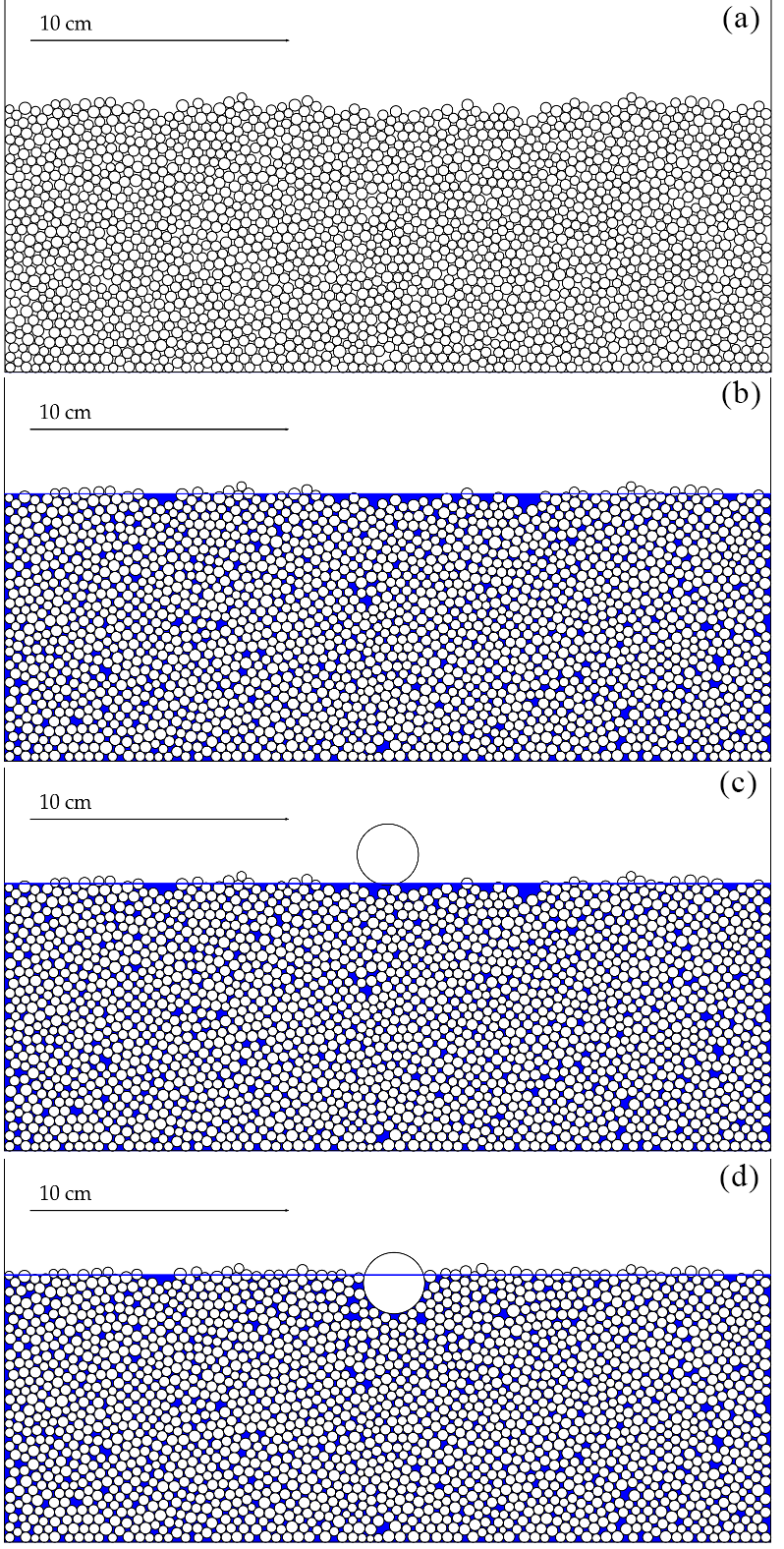}
        \caption
	[Creation of the initial state]{The different steps to create the initial state:
	(a) First a granular medium of 2000 particles is created.
	(b) Next, water is added in the porous volume (blue part).
	(c) Further, an intruder (of radius 12 mm here) is hung on the top of the medium and released.
	(d) Eventually the intruder relaxes to a state of mechanical equilibrium with the medium.}
        \label{fig:simu}
\end{center}
\end{figure}

We create horizontal lines of particles by making $100N_{MAX}$ particles with random horizontal positions,
at exactly $2\,r_{\text{mean}}$ above the lowest altitude free of particles, and then remove overlapping ones.
The particles radii follow a normal law centered around $r_{\text{mean}}$ with a standard deviation of $8\%$
of $r_{\text{mean}}$.
The line is set free to fall and reach mechanical equilibrium.
This procedure goes on until the desired number of particles is reached.
The final porosity is between 0.196 and 0.199 which is characteristic of a random loose pack for a 2D granular medium \cite{Aharonov1999,Donev2004}.
Once this initial soil skeleton is in place,
we measure the final height of the granular medium by computing the mean of the vertical position
of the last layer of beads.
We fix the water level at that height in order to have a saturated medium.
This configuration of granular media is representative of a soil saturated with water
which is the typical soil where liquefaction and quicksands occur
\cite{ambraseys1969liquefaction,Wang2010,Khaldoun2005}.
We fix the height of an intruder at the surface of this new saturated medium,
and release it.
The size of the intruder is chosen as 6 times the linear size of the small particles, 
so that it is significantly larger than them, and remains small enough compared to the size of the box, to avoid finite size effects.
The parameters used for the simulations presented in this paper
are summarized on Table \ref{tab:ex}.
A representation of the different steps to create the final medium
is given in Fig.~\ref{fig:simu},
where $r_{\text{mean}}=2 \mathrm{\ mm}$ and $w_{BOX} = 30 \mathrm{\ cm}$.
%*** MENKA: maybe crop the first image and replace it with an image where a line of grains is added.

The next stage is the main part of the simulations.
Here we impose a horizontal movement on the two lateral walls of the box.
Both sides move synchronously, following a sinusoid.
We record the positions and the velocities of all particles every hundred timesteps.

\begin{table}
\begin{ruledtabular}
\begin{tabular}{lrl}
	Radius of particles  & $r_{\text{mean}} \pm 8\%$ & $2 \pm 0.16 \mathrm{\ mm} $\\
	Density of particles & $\dso$&  $1.05 \mathrm{\ g\,L^{-1}}$\\
	Elastic constant (during shocks) & $k$& $20000\mathrm{\ kg\,s^{-2}}$\\
	Viscosity constant (during shocks)  & $\nu$ & $0.3\mathrm{\ N\,s\,m^{-1}}$\\
	Friction coefficient & $\mu$& 0.6 \\
	Cohesion & $c$& 0.0\\
	Number of particles & $N$& 2000\\
	Time step & $dt$& $10^{-6} \mathrm{\ s}$\\
	Box size & $w_{BOX}\times L$& $ 30 \mathrm{\ cm} \times 30  \mathrm{\ cm}$\\
	Radius of the intruder & $\rin$ & $12 \mathrm{\ mm} $\\
	Density of the intruder & $\din$ & $1.0 \mathrm{\ g\,L^{-1}}$\\
\end{tabular}
\end{ruledtabular}
\caption{\label{tab:ex}Parameters used for the simulations.}
\end{table}

%%%%%%%%%%%%%%%%%%%%%%%%%%%%%%%%%%%%%%%%%%%%%%%%%%%%%%%%%%%%%%%%%%%%%%%%%%%%%%
%%%%%%%%%%%%%%%%%%%%%%%%%%%%%%%%%%%%%%%%%%%%%%%%%%%%%%%%%%%%%%%%%%%%%%%%%%%%%%
%%%%%%%%%%%		THRESHOLDS DEMILITING FLOW TYPES 	%%%%%%%%%%%%%%
\subsection{Thresholds delimiting flow types}
\label{sec:thresholds}
\subsubsection{Variables which quantify the intruder movement}

In both the computer simulations and experiments,
we record the temporal evolution of  the height of the intruder.
From this height and the height of the granular medium
we compute the immersed depth of the intruder $h(t)$ as 
the distance between the surface of the medium and the bottom of the intruder.
The immersed volume of the intruder $\vinim$ is related in 3D to $h(t)$ by the following relation:
\begin{equation}
\vinim(t) = \frac{\pi}{3}(3rh^2(t) - h^3(t))
\label{eq:vim}
\end{equation}
To compare our results with other sizes or shapes of intruder,
we will express our computation in term of $\vnorm(t)$,
the intruder's emerged volume $\vinem(t)$ normalized by its initial emerged volume $\vinem(0)$
and its final emerged volume $\vinem(\infty)$.
$\vnorm(t)$ is defined as follows:
$$\vnorm(t) = \frac{\vinem(t) - \vinem(\infty)}{\vinem(0) - \vinem(\infty)}$$
$$ = \frac{\vin-\vinim(t) - \vin+\vinim(\infty)}{\vin-\vinim(0) - \vin+\vinim(\infty)}$$
%%\begin{equation}
%%	\vnorm(t) = \frac{\vinim(\infty)- 3rh^2(t) + h^3(t)}{\vinim(\infty)- \vinim(0)}
%%\label{eq:nev}
\begin{equation}
	 = \frac{\vinim(\infty)- \pi/3(3rh^2(t) - h^3(t))}{\vinim(\infty)- \vinim(0)}
\label{eq:nev}
%% einat - was there a typo here? should be multiplied by pi/3? i corrected. 
\end{equation}
The term $\vinim(0)$ is the immersed volume of the intruder during the initial state,
when it is at rest on the medium.
Here we assume $\vinim(\infty)$ to be the theoretical isostatic immersed volume of the intruder
$\vinim^{\text{equilibrium}}$,
computed for an immersion in a fluid of density $\deff$, according to Eq.~(\ref{eq:vinimeq}).
For all simulations and experiments $\vnorm(t)$ starts at 1 and decreases
as the intruder sinks.
If the intruder reaches the isostatic equilibrium given in Eq.~(\ref{eq:vinimeq}), then $\vnorm(t)$ reaches 0.

\begin{figure}[htbp]
\begin{center}
		\includegraphics[width=8cm]{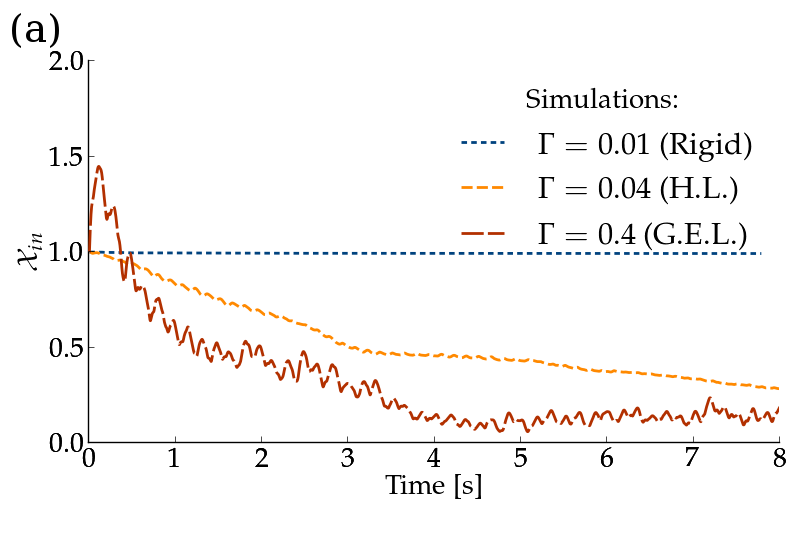}
	\includegraphics[width=8cm]{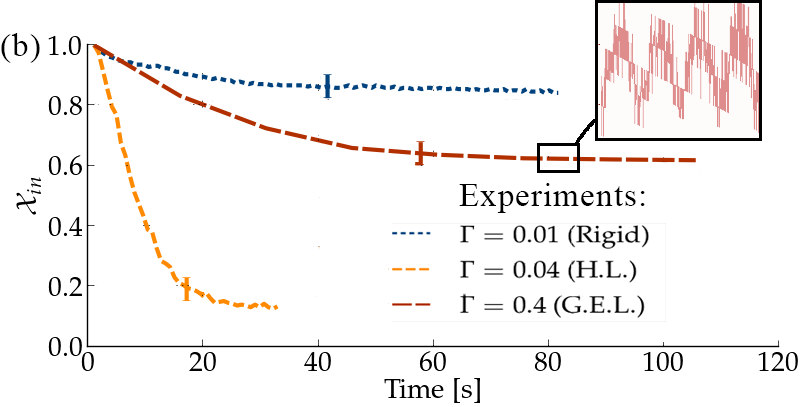}
	\caption
{Normalized emerged volume of the intruder $\vnorm$ as a 
function of time for the three different regimes described in 
the first section, in the case of (a) simulations and (b)
experiments. The experiments and simulations are done in a saturated granular medium.
For simulations the frequencies used are $12 \mathrm{\ Hz}$ for the rigid and G.E.L. cases, and $7 \mathrm{\ Hz}$ for the H.L. case.
For experiments the frequencies used are $0.8 \mathrm{\ Hz}$ for the rigid case, $1.6 \mathrm{\ Hz}$ for the H.L. case
and $50 \mathrm{\ Hz}$ for the G.E.L. case.
}
	\label{im:3behav}
\end{center}
\end{figure}

We show on Fig.~\ref{im:3behav} the evolution of $\vnorm$ for three simulations and three experiments
showing the typical behaviors of the three deformation regimes,
rigid, H.L. and G.E.L..
In the rigid cases (blue curves), $\vnorm$ stays close to 1.
A small descent exists anyway but it can be attributed to the compaction of the medium.
In the H.L. cases (orange curves), $\vnorm$ slowly decreases
from 1 to a final value between 0.2 and 0.
The G.E.L. behavior is characterised by an irregular descent of the intruder,
with relatively high fluctuations around the main trend of the curve, continuously  perturbing the equilibrium state.
%Eventually the global excitation behavior is different for simulations and experiments.
%In both cases the intruder does not follow a regular descent toward 
%the isostatic state.
%It sinks slightly in the experiments but does not approach the isostatic position. 
%It sinks fast in simulations and oscillate around a deep position.
%The numerical granular medium is decompacted and all the particles are in movement.
%%%RENAUD: ON DEVRAIt PRECISER UNE INTERPRETATION DE LA DIFFERENCE. TAILLE DES GRAINS?
During experiments, the use of the fast camera is required
to see that the intruder is continuously oscillating with high frequencies during G.E.L. states (see zoom on Fig.~\ref{im:3behav}).
Even without the fast camera, 
one can observe with the naked eye that G.E.L. states exhibit
convection cells characterized by particles at the surface going from the middle of the box toward its sides.

In the previous well-selected cases the behavioral regimes of the system were obvious. 
Nevertheless this is not always the case, and especially not when the excitation is on the limit between two regimes. 
Hence, we need to specify quantitative criteria to automatically differentiate the three regimes among all the experiments and simulations.
In the following paragraph we precise
the exact criteria and thresholds that we use in practice.

\subsubsection{Thresholds between rigid and heterogeneous liquefaction (H.L.)  states}
\label{thresholdrhl}
The medium is categorised to be in the rigid state when the intruder does not move significantly downwards.
Actually, the intruder usually sinks slightly because the medium compacts during  shaking.
In simulations, the medium compacts less, possibly due to the fact that the movement takes place in 2D,
and there are less degrees of freedom for rearrangements in 2D than in three dimensions (3D).
According to the observations
we categorize as rigid the experiments where  $\vnorm$ decreases in total less than $10\%$ from its initial value,
and in the simulations where it decreases less than $5\%$ from its initial value.
The exact choice of these threshold values does not affect significantly the phase diagram we will obtain.

\subsubsection{Thresholds between heterogeneous liquefaction (H.L.) and global excitation liquefaction (G.E.L.) states}
\label{thresholdhlgel}
When $\vnorm$ decreases by more than $10\%$  during experiments,
or more than $5\%$ during simulations, we categorize the medium state either as the
heterogeneous liquefaction (H.L.) case, or as the global excitation liquefaction (G.E.L.) case.
The distinction between these two cases is done as follows:
From a macromechanical point of view, the G.E.L. state starts when the intruder
keeps oscillating around a final position without reaching a final equilibrium.
Depending on the frequency these oscillations can be small and fast or large and slow.
A good criterion to determine the category is to base the distinction on the measure of the acceleration of these oscillations.
This method allows to catch the G.E.L. cases at both small and high frequencies.
When the standard deviation of the acceleration of the intruder
is greater than $0.6 \mathrm{\ m\,s^{-2}}$,
the simulations and experiments are classified as G.E.L. cases.

%%%%%%%%%%%%%%%%%%%%%%%%%%%%%%%%%%%%%%%%%%%%%%%%%%%%%%%%%%%%%%%%%%%%%%%%%%%%%%%%%%%
%%%%%%%%%%%%%%%%%%%%%%%%%%%%%%%%%%%%%%%%%%%%%%%%%%%%%%%%%%%%%%%%%%%%%%%%%%%%%%%%%%%
%%%%%%%%%%%%%%%%%%%%%%%%%%%%%%%%%%%%%%%%%%%%%%%%%%%%%%%%%%%%%%%%%%%%%%%%%%%%%%%%%%%
%%%%%%%%%%%%%%%%%%%%%%%%%%%%%%%%%%%%%%%%%%%%%%%%%%%%%%%%%%%%%%%%%%%%%%%%%%%%%%%%%%%
%%%%%%%%%%%%%%%%%		RESULTS		%%%%%%%%%%%%%%%%%%%%%%%%%%%%%%%%%%%
\section{Results}
\subsection{Water influence on soil liquefaction}

The first interesting result is the strong effect of the presence of water on the behavior displayed by the medium.
To highlight the role of water in soil liquefaction, we compare the behavior of saturated and dry granular media.
We first focus on laboratory experiments.
\begin{figure}[htbp]
\begin{center}
	\includegraphics[width=7cm]{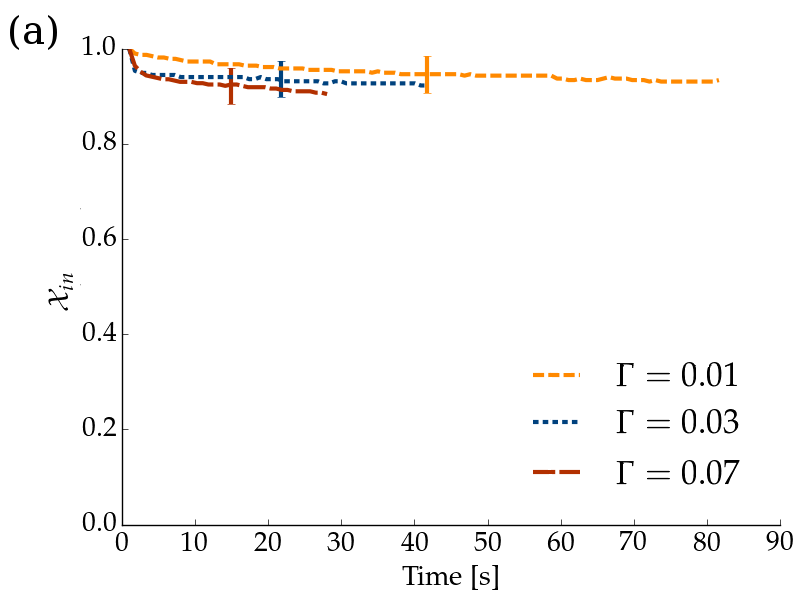}
\end{center}
\begin{center}
	\includegraphics[width=7cm]{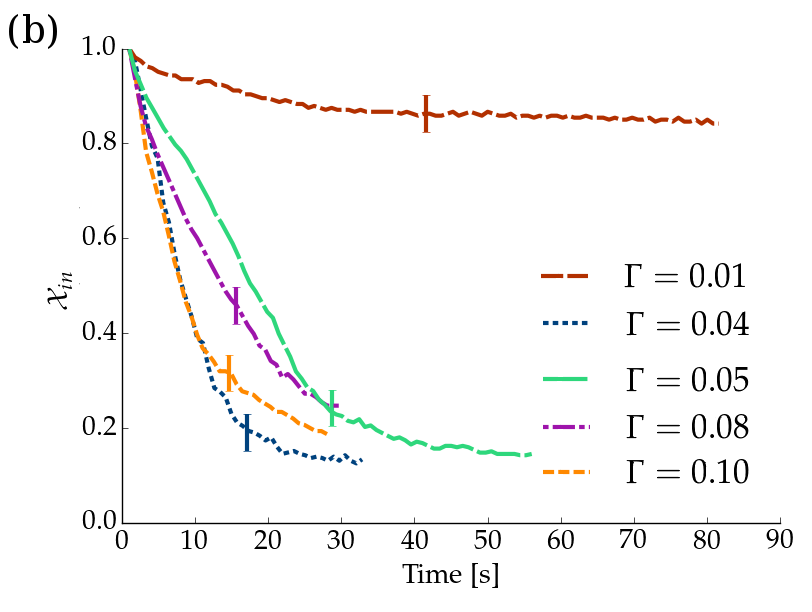}
	\caption[]{
	Normalized emerged volume $\chi_{in}(t)$ as function of time during several experiments (a) in a \emph{dry} medium and
	(b) in a \emph{saturated} medium.
	For the dry cases the horizontal shaking has an amplitude of 7 mm and
	its frequency varies from 1.5 to 3.5 Hz.
	For the saturated the shaking has an amplitude of 3.5 mm and
	its frequency varies from 0.8 to 2.5 Hz.}
	\label{im:expedry}
\end{center}
\end{figure}
We compare $\vnorm$ for experimental media fully saturated to the top of the grains, (Fig.~\ref{im:expedry}~(b)),
and dry experimental media (Fig.~\ref{im:expedry}(a)), shaken by the
same force. 
For the saturated medium the transition 
between the rigid behavior and heterogenous liquefaction is obvious: 
the intruder remains on the surface
for the lowest acceleration ($\Gamma = 0.01$, rigid), 
but for accelerations larger than $\Gamma \geq 0.04$,
the intruder sinks quickly into the medium (liquefaction).
On the contrary, in the dry case, for any acceleration between 
$\Gamma = 0.01$ to $\Gamma = 0.07$ the intruder does not sink. 
We did not observe the G.E.L. behavior in neither the dry or saturated cases, since the
results shown in Fig.~\ref{im:expedry} were obtained using the setup of
Fig.~\ref{fig:setup} A, and the accelerations reached with this setup were too low. G.E.L. was observed for different cases where $\Gamma>1$ both in dry and saturated media, using the setup of Fig.~\ref{fig:setup} B.

\begin{figure}[htbp]
\begin{center}
	\includegraphics[width=7cm]{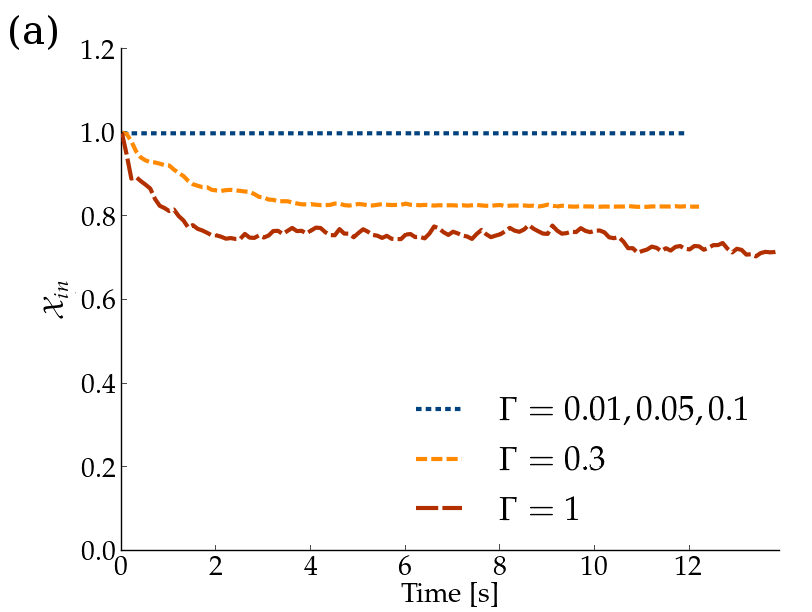}
\end{center}
\begin{center}
	\includegraphics[width=7cm]{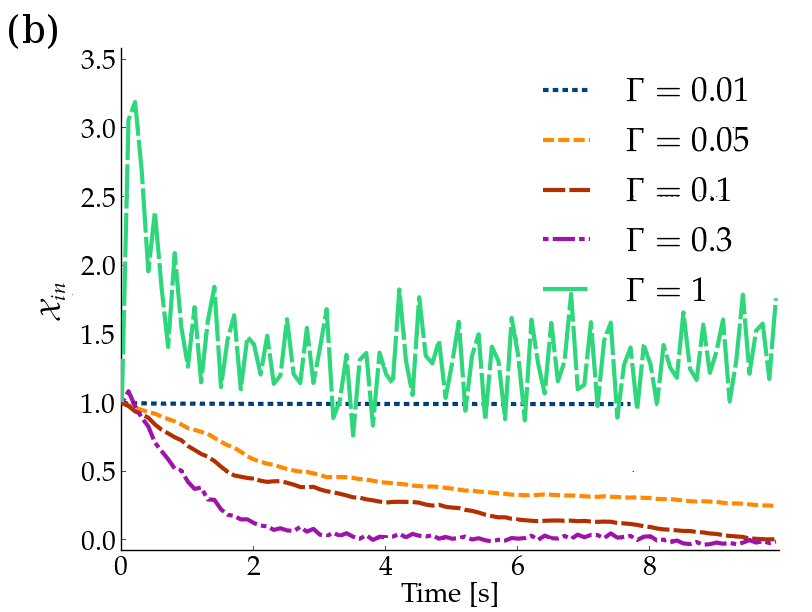}
	\caption{
	Normalized emerged volume $\chi_{in}(t)$ as function of time during several simulations ran in (a): a \emph{dry} medium,
	and (b): a \emph{saturated} medium.
	The horizontal shaking has a frequency of 12 Hz for the four first curves,
	its amplitude varies from 0.02 mm to 0.5 mm.
	For the curve $\Gamma = 1$ we used a frequency of 7 Hz and an amplitude of 5 mm.}
	\label{im:simu}
\end{center}
\end{figure}
The same feature is observed in simulations, as shown in Fig.~\ref{im:simu}.
The medium remains rigid for $\Gamma = 0.01$ in both cases of dry and saturated media. 
An important sinking due to H.L. is observed for $\Gamma \in [0.05; 0.1]$
in the case of saturated media only.
Eventually the G.E.L. behavior can be observed on Fig.~\ref{im:simu} for $\Gamma \in [0.3; 1]$.
%%Both for dry cases and saturated cases, the intruder displays a large permanent movement,
%%following the waves observed at the surface of the saturated medium, 
%%in response to the shaking of the box.
%% einat -the above  sentence is confusing. 
While for the saturated medium, Fig.~\ref{im:simu}(b) shows
the transition between the three described regimes: rigid, H.L. and G.E.L. , the case of a dry granular medium, Fig.~\ref{im:simu}(a), 
shows a direct transition from the rigid case 
($\Gamma \in [0.01; 0.1]$) to G.E.L. ($\Gamma \geqq 0.3$), 
without passing through the H.L. regime.
There is no value of $\Gamma$ where the intruder descends further beyond $5\%$ (for $\chi_{in}$)
and where the standard deviation of its acceleration remains lower than $0.6 \mathrm{\ m\,s^{-2}}$ at the same time.

These initial results confirm that the presence of water strongly promotes liquefaction, and is required to
produce liquefaction at moderate shaking accelerations \cite{Clement2016}. They are qualitatively consistent with the predictions of the simple model
summed up in Eqs. ~(\ref{eq:rig}, \ref{eq:liq}) and (\ref{eq:flu}) for saturated cases and 
in Eq.~(\ref{eq:thresholddry}) for dry cases.
Both in the experiments and in the simulations,
the shaken granular medium liquefies easily when water is added to the granular medium.
In our simulations the presence of water is represented by local buoyancy without any compressibility or viscosity effects.
Note that the pore pressure is thus increased in the saturated case with respect to the dry case, since it is hydrostatic, but it is not increased further during the simulations (it always stays hydrostatic): the rheology change of the saturated granular media is thus not attributabe to dynamic pore pressure rise.
Liquefaction is triggered in our experiments and simulations by external shaking,
with a top drained boundary condition where the water is not confined
-- i.e. where water can flow in and out of the surface.
Inside the granular system the water acts solely through a buoyancy force
and reduces the effective weight of the particles.
The effective stress is reduced and consequently,
grains can slide more easily on each other in presence of these buoyancy forces. 
The sliding motion allows liquefied deformation of the granular media.   
The importance of buoyancy is in enlarging the range for this sliding onset,
allowing it to  occur under rather low accelerations,
and, more importantly,
in inducing a crucial difference between the grains and the intruder:
the later being only partially immersed,
the intruder-grains contacts are stronger than grain-grain contacts.
%RENAUD: JE PENSE QU'ON A BESOIN D'UN COMMENTAIRE DISANT QUE DANS LES CAS DE LIQUEF OBSERVEE SUR LE TERRAIN, EN GENERAL, ON NE VOIT PAS DE MOUVEMENT DE TERRAIN AUTOUR D'UN BATIMENT QUI SUBSIDE, MAIS PLUS LOIN: CA COLLE BIEN A NOTRE MODELE, EXPS ET SIMULS.

%%%%%%%%%%%%%%%%%%%%%%%%%%%%%%%%%%%%%%%%%%%%%%%%%%%%%%%%%%%%%%%%%%%%%%%%%%%%%%%%%%%
\subsection{Micromechanical point of view}
%%%%%%%%%%%%%%%%%%%%%%%%%%%%%%%%%%%%%%%%%%%%%%%%%%%%%%%%%%%%%%%%%%%%%%%%%%%%%%%%%%%

\begin{figure}[htbp]
\begin{center}
\includegraphics[width=8cm]{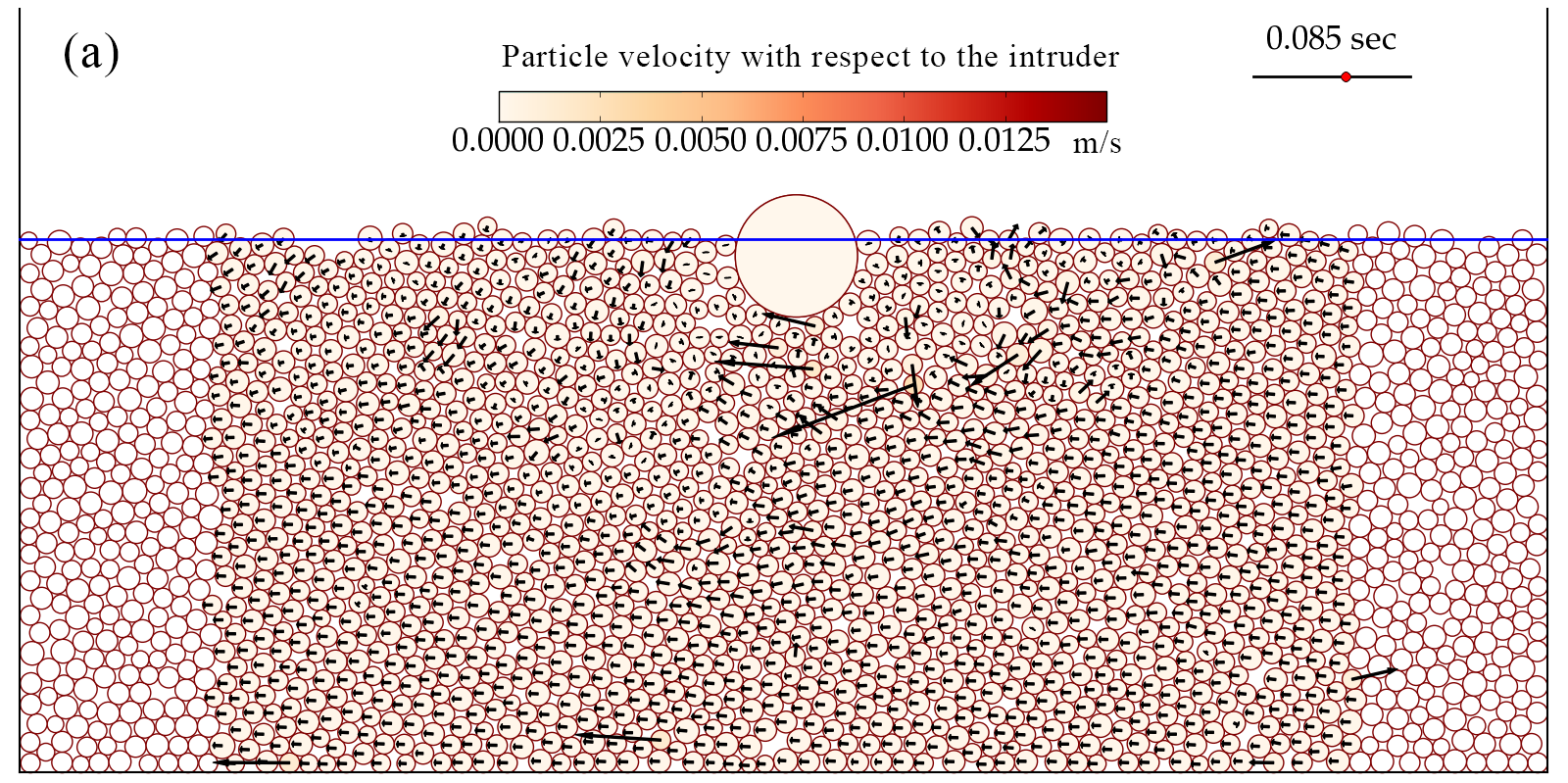}
\includegraphics[width=8cm]{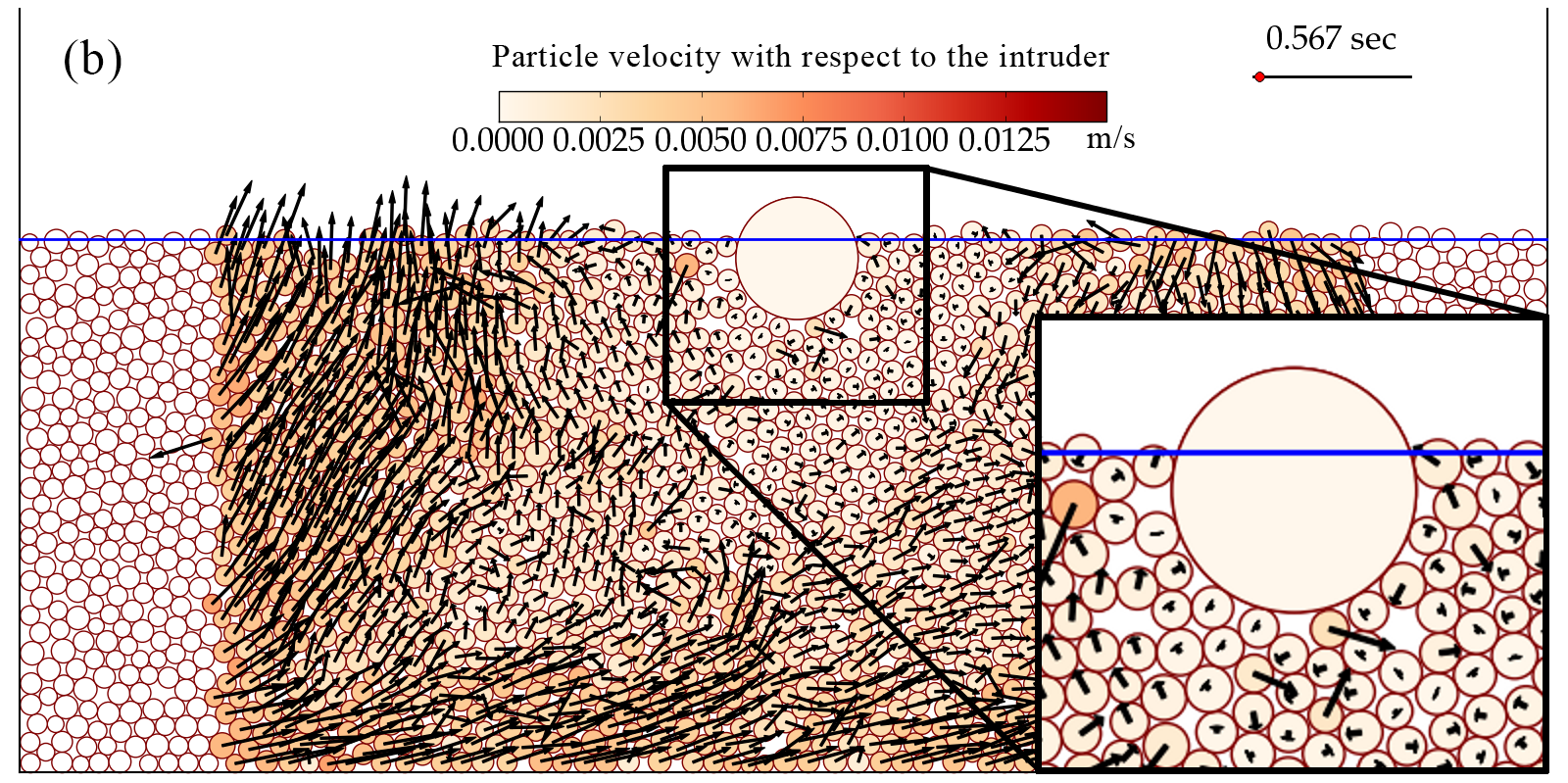}
\includegraphics[width=8cm]{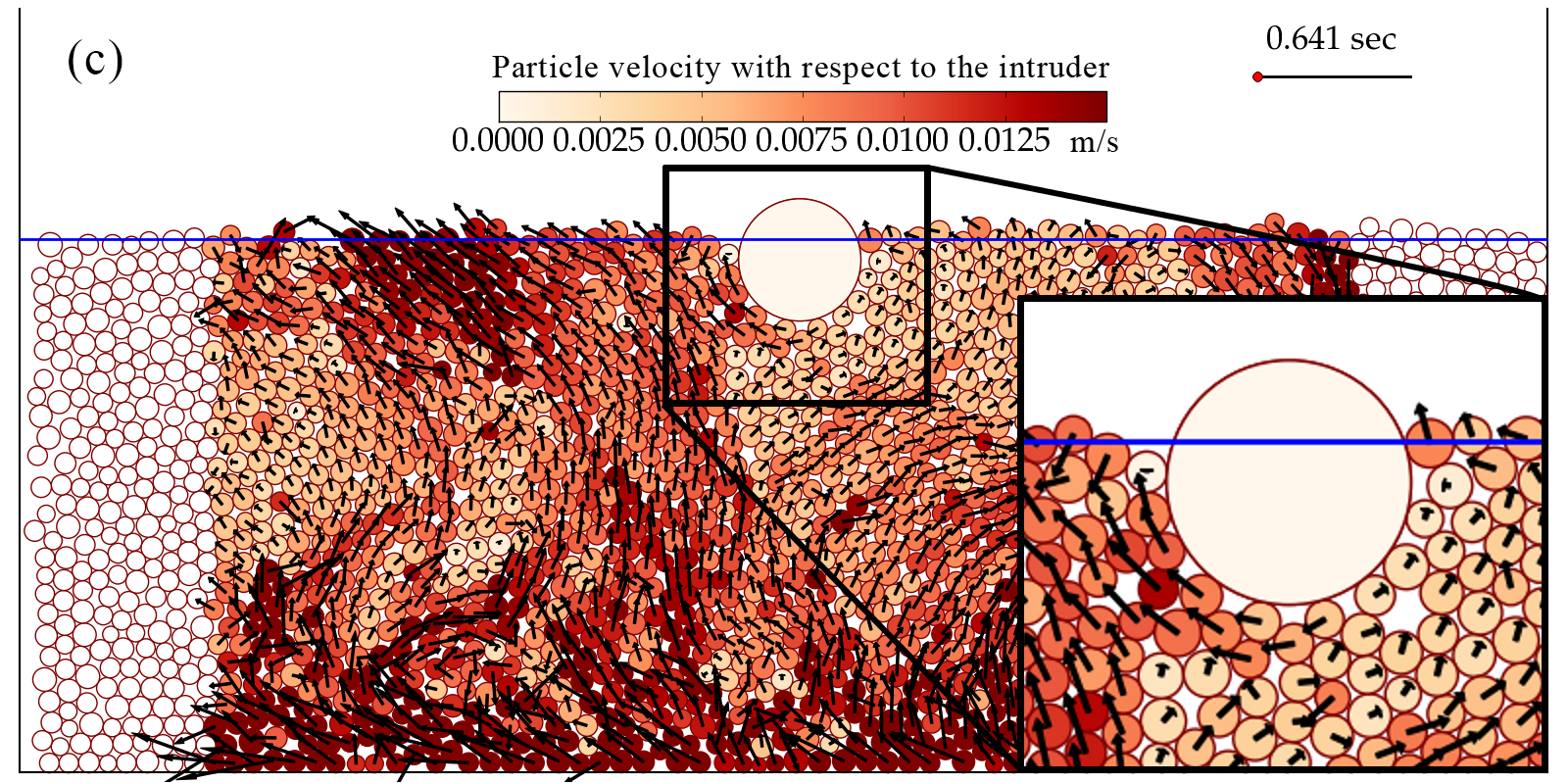}
\end{center}
\caption[]{Micromechanics of (a) a rigid case, (b)
a H.L. case and (c) a G.E.L. case: snapshots of the system configuration and velocity field.
The arrows represent the deviation velocity of the particles (velocity with respect to the intruder) and the color of the particles represent this deviation velocity norm in $\mathrm{\ m\,s^{-1}}$.
The blue line is the water level. The straight dark line on the top right represents the trajectory of the moving box (arbitrary scale), and the red dot the position of the box in the cycle at the time of the snapshot - the time elapsed since the start of the shaking is indicated above the line.
The deviation velocity is not represented for particles close to the borders, since their
behavior is dominated by the shocks agains the walls and not the liquefaction process.}
	\label{im:micro}
\end{figure}

To better understand what governs the sinking of the intruder during our experiments and simulations,
we focus on the deformations inside the granular medium and will here adopt a micromechanical point of view.
The simulations allow us to follow in detail every particle inside the medium,
and to investigate the physics of liquefaction.
%After having looked at the velocity vector field
%we comput the norm of the velocity of each particles minus the velocity of the intruder,
%we call this value the deviation velocity, hint the deviation from the intruder velocity.
The explanation of liquefaction proposed in subsection \ref{section:problematic} is based
on the possibility or not for the intruder to slide on the particles beneath it.
Hence, verifying its validity necessitates considering the relative velocity
between the intruder and the grains beneath. 
For this purpose we define the ``deviation velocity" as the velocity of the grains in the 
reference frame of the intruder, i.e. the grain velocities minus the velocity of the intruder.
The deviation velocity of the particles is represented in Figs.~\ref{im:micro}, for three simulations carried out in a saturated medium shaken at a frequency of 12 Hz, at three different values of the normalized PGA, $\Gamma$.
Each snapshot shows the state of the system at a given time, with the arrows pointing in the
direction of the deviation velocity, and the particles' color corresponding to the deviation
velocity module. 

One can observe that for $\Gamma=0.01$, i.e. the rigid state
shown on Fig.~\ref{im:micro}(a),
the deviation velocity is almost zero in the whole medium,
and the intruder follows the movement of the surrounding particles.
Every particle follows the imposed movement of the box, and the medium does not deform.
Only few particles move or roll because of local compaction, or because they are free at the
surface of the medium.
For $\Gamma=0.28$ the system is in the H.L. case, where we expect
that the intruder cannot slide on the particles beneath it ($\Gamma<\mu$), but particles can slide
on each other ($\Gamma>\frac{\rho_S-\rho_W}{\rho_S}\mu$). 
Hence, the object is not directly sliding on the surrounding particles
and stays fixed to them,
while the whole granular medium, far from the intruder,
is able to undergo sliding and to deform easily - 
leading to the subsidence of the intruder and of the granular medium under it.
The analysis of the deviation velocity shows that this is indeed the case: 
Figure \ref{im:micro}(b)
shows that the deviation velocity around the intruder is weak (under $2.5 \mathrm{\ mm\,s^{-1}}$). 
Hence, there is no sliding between these particles and the intruder - the intruder and its neighbours move together.
However, the particles further away from the intruder (a few intruder diameters) are in motion with respect to the intruder.
This shows that the medium is rearranging, and as a result, the intruder and its
surrounding particles sink as a whole with respect to this farfield.
%The deviation velocity field can be described like a current made by the particles surrounding the intruder
%into the whole medium, while the particles further away experience a reduced solid normal stress and sliding, which allows the intruder to subside, i.e. to sink slowly.
%einat- the above sentence is a re[eat of what is said above and below. 

Finally, for $\Gamma=0.7$ (Fig.~\ref{im:micro}(c)) the system is in the G.E.L. regime, 
which can be charachterized by the sliding of the intruder on the surrounding particles ($\Gamma>\mu$). 
In this case the velocity deviation inside the G.E.L. media
is roughly 10 times larger than in the H.L. case.
Under the intruder the  velocity deviation of the particles is between 
$2.5 \mathrm{\ mm\,s^{-1}}$ and $10 \mathrm{\ mm\,s^{-1}}$ (Fig.~\ref{im:micro}(c)).
This non-zero relative velocity between the intruder and its neighbors shows that, 
indeed, the intruder slides on the particles beneath it, 
hence it behaves like any other particle of the medium.
It is during this behavior that we may observe convection cells which drag the particles along cells
connecting the bottom and the top of the medium. This is the case in the example of Fig.~\ref{im:micro}(c).
%This behavior is also characterized by  convection cells
%which drag the particles along cells connecting the bottom and the top of the medium.
%%%%%%%%%%%%%%%%%%%%%%%%%%%%%%%%%%%%%%%%%%%%%%%%%%%%%%%%%%%%%%%%%%%%%%%%%%%%%%%%%%%
\subsection{Phase diagram controlling liquefaction occurence and type}
%%%%%%%%%%%%%%%%%%%%%%%%%%%%%%%%%%%%%%%%%%%%%%%%%%%%%%%%%%%%%%%%%%%%%%%%%%%%%%%%%%%

Three general types of behavior have been identified and analyzed from both
a macromechanical and a micromechanical point of view.
We will now examine under which conditions these different behaviors occur,
derive a phase diagram as function of the control parameters, and check using experiments and simulations
the theory derived in subsection \ref{section:problematic} for the transition between these three behaviors.
For this purpose, we make a systematic series of experiments and simulations
at various frequencies and amplitudes.
\begin{figure}
\begin{center}
   \includegraphics[width=8cm]{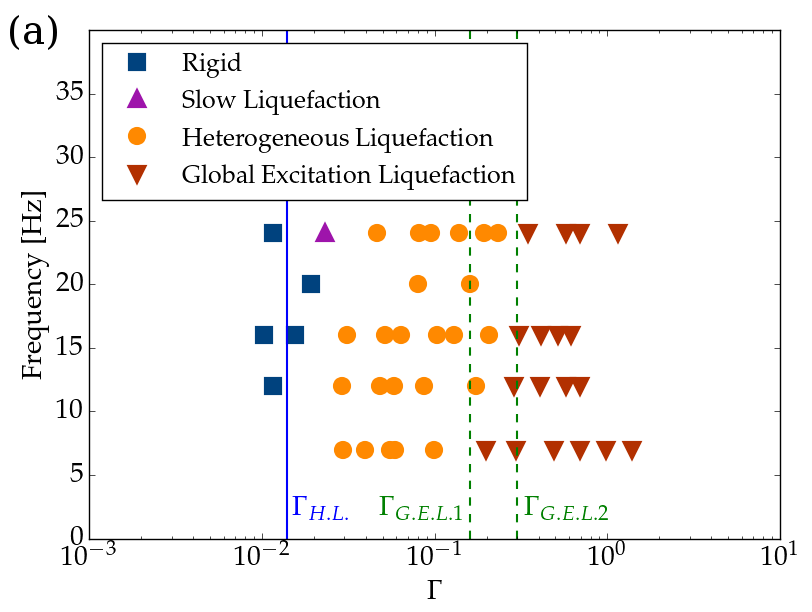}
	%\label{fig:phases_diagsimu}
 %  \\A) Simulations
\end{center}
%\end{figure}
%\begin{figure}
\begin{center}
   \includegraphics[width=8cm]{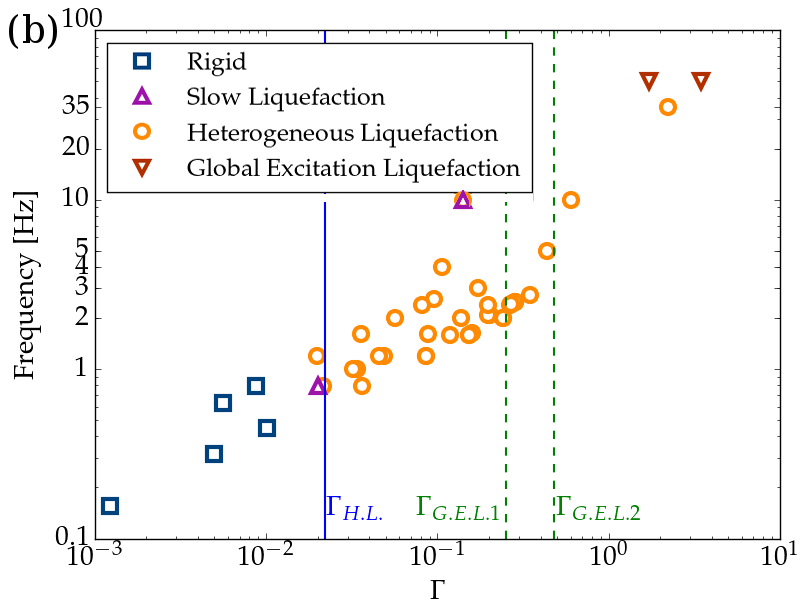}
 %  \\B) Experiments
\end{center}
 \caption[Phase diagram]{Phase diagram of the numerical simulations (a) and of the experiments (b).
	Each simulation or experiment is plotted according 
	to its reduced acceleration $\Gamma$ and its frequency.
	The behaviors are determined following the thresholds defined in subsections \ref{thresholdrhl} and \ref{thresholdhlgel}.
	Squares correspond to an observed rigid behavior, circles and triangles pointing up to H.L. behavior,
	and triangles pointing down to G.E.L. behavior.
	The boundaries theoretically derived in section \ref{section:problematic} between these three regimes
	are the vertical lines: $\Gamma=\gammagm$ for the rigid/H.L. boundary,
	and $\Gamma=\gammab$$_1$ or $\Gamma=\gammab$$_2$  are two
	possibilities for the H.L./G.E.L. boundary.
	These boundaries match well with the observed symbol changes, both in experiments and in simulations.}
\label{fig:phases_diagexpe}
\end{figure}
The frequencies range from $7 \mathrm{\ Hz}$ to $24 \mathrm{\ Hz}$ for simulations and
from $0.15 \mathrm{\ Hz}$ to $50 \mathrm{\ Hz}$ for experiments and
the acceleration range from $0.01 \mathrm{\ m\,s^{-2}}$ to $1 \mathrm{\ m\,s^{-2}}$ for simulations and
from $0.001 \mathrm{\ m\,s^{-2}}$ to $4 \mathrm{\ m\,s^{-2}}$ for experiments.
%$10^{-2} \mathrm{\ m\,s^{-2}}$
We link each experimental and simulation run to one of the three behaviors,
rigid, H.L. or G.E.L., according to the categorization criteria explained in subsections
\ref{thresholdrhl} and \ref{thresholdhlgel}.
On the phase diagram, Fig.~\ref{fig:phases_diagexpe}, we show the results of this systematic study for the numerical simulations (a) and the experiments (b).
$\Gamma$ is represented on the horizontal axis and the frequency of shaking is on the vertical axis.
Each simulation and experiment is plotted with a particular symbol representing the associated behavior:
blue squares for rigid states, orange discs for H.L. states and red triangles pointing down for G.E.L. states,
according to the thresholds defined in section \ref{sec:thresholds}.
A particular liquefaction behavior, represented by purple triangles pointing up and called slow liquefaction,
will be further discussed in the next section.
It is related to a few experiments and simulations which don't follow the same master curve as all
other experiments and simulations.
%RENAUD: VU QUE TU NE DEFINIS PAS DANS CETTE PHRASE LA SLOW LIQUEF, NI LA MASTER CURVE, LE MIEUX EST DE LE GARDER POUR PLUS TARD, QUAND ON PARLE DE LA VITESSE DE PENETRATION
As predicted by the theory and confirmed by  figure \ref{fig:phases_diagexpe} both for simulations and experiments,
the main control parameter determining whether liquefaction happens, and what type of liquefaction,
is the value of the normalized peak ground acceleration $\Gamma$.

Two possibilities are shown for $\gammab$: $\gammab$$_2$ corresponding to the theoretical sliding threshold of
an initialy emerged intruder (Eq.~(\ref{eq:threshold}) of section \ref{section:problematic}) and
$\gammab$$_1$ corresponding to the sliding threshold of an intruder which is initially partially immersed.
$\gammab$$_1$ is calculated as $\gammab$$_1 = \mu (1-\frac{\vinim(0) \dwa}{\vin \din})$, with the initial immersed
volume $\vinim(0)$ of 50\% for the simulations and 12\% for experiments.
%On this figure we computed two possibilities for the sliding theoretical boundary:
%$\gammab$$_1=$$\gammab$ is the one described and computed on section \ref{section:problematic}, Eq.~(\ref{eq:threshold}).
%For this theoretical boundary, based on the acceleration at which
%the intruder starts to slide on the particles under it,
%we assumed the intruder to be initialy entirely above the medium.
%However our simulations start with an intruder half immersed in the medium as initial state,
%which comes from mechanical equilibrium without shaking.
%As for the experiments, between 5\% and 20\% of the volume of the intruder is initialy immersed before shaking.
%Thus $\gammab$$_2$ is the theoretical boundary with this corrected initial intruder position,
%$\gammab$$_2 = \mu (1-\frac{\vinim(0) \dwa}{\vin \din}).$
$\gammab$$_1$ is smaller than $\gammab$$_2$ because the buoyancy applied on the intruder immersed volume
reduces the acceleration needed to make it slide on the particles underneath.\\

Let us examine, from Fig.~\ref{fig:phases_diagexpe},
the deviation between the phase boundaries derived experimentally or numerically,
and those obtained with the simple analytical model,
considering $\gammab$$_1$ the threshold corrected for the initial partial immersion of the intruder.
The boundary between rigid and H.L. state is very well reproduced
by both the simulations and the experiments.
%With the experiments it appears that the medium liquefies at smaller acceleration than the predicted one.
%This can be due to the first setup we used for experiments at low frequency of shaking.
%Indeed because the engine was moving step by step we noticed 
%extra high frequencies crossing the medium visible thanks to fast surface waves on the water.
%These high frequencies could be the source of non-controled high acceleration and trigger liquefaction.
%However they remain weak and the observed limit for onset liquefaction is very close to the theoretical one.
Concerning the boundary between H.L. and G.E.L.
the phase diagram of the simulations shows again a very good fit beween theory and experiments.
For the experiments, the setup limitations do not allow too many experiments at very large accelerations.
A dispersion of the behavior results is observed,
with a gradual transition from H.L. to G.E.L..
The transition nonetheless happens at a central value around the one predicted by theory.
At first order, the two theoretical boundaries $\gammagm$ and $\gammab$ capture
very well the location of the different behaviors observed with numerical simulations and experiments.
The similarity found in the results between simulations, experiments and theory are thus satisfactory, and validate
the explanation proposed for the physics of liquefaction.

%ON PEUT AJOUTER LE CAS DE KHALDOUN ET DIRE QUE CA EXPLIQUE AUSSI LES OBSERVATIONS DE SABLE MOUVANT.

%CE SERAIT BIEN D'AVOIR AUSSI UNE PETITE SECTION SUR LA PROFONDEUR FINALE ATTEINTE, ET La COMPARAISON A L'ISOSTATIQUE, POUR LEs EXPS ET LES SIMS.
\subsection{Comparison of the final position and the isostatic position}

\begin{figure}[htbp]
\begin{center}
\includegraphics[width=8cm]{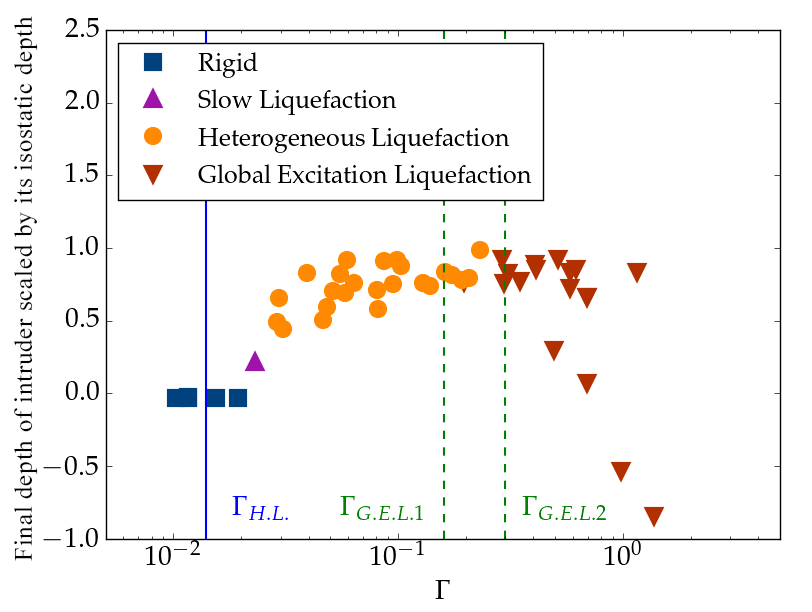}
\end{center}
\caption[]{Final depth of the intruder scaled by its isostatic depth, $h(\infty)/h_{ISO}$,
as function of the dimensionless imposed peak ground acceleration $\Gamma$.
The markers correspond to the behavior of the medium, rigid, slow liquefaction, H.L. or G.E.L.} 
\label{im:final_position}
\end{figure}

In the previous paragraph we show that the behavior of the intruder above the shaken granular media
can be sorted into three cases according to the imposed acceleration,
as we expected given the theoretical analysis.
We will now study what is the final position of the intruder during H.L.
and the velocity at which it approaches it.
During the H.L. regime, as a first approximation, if the vertical friction forces on the intruder average to zero after penetration, the intruder will approximately approach the 
theoretical isostatic position dictated by its weight and the buoyancy (section \ref{theoretical_final_position}).
The experimental and numerical setup enables to test whether this approximation holds or fails, by measuring precisely the final vertical position of the intruder
for a comparison with the isostatic position. 
For each simulation the ratio of the final position of the intruder to its isostatic position, $h(\infty)/h_{ISO}$, is represented on Fig.~\ref{im:final_position}. Different symbols correspond to different behaviors observed fo the simulations: rigid, slow liquefaction, H.L. or G.E.L..
%The slow liquefaction behavior, represented by purple triangles pointing up is a particular case of liquefaction
%happening in a narrow range of acceleration around the threshold between rigid behavior and H.L.
%which will be further discussed in the last section.
We see that when $\Gamma$ rises, the ratio rises towards 1, i.e. the
intruder approaches its isostatic depth.
The ratio is 0.2 for the slow liquefaction case, and 
between 0.5 and 1 for most H.L. cases.
When $\Gamma$ exceeds 0.2 to 0.3 
the media displays G.E.L. behavior, and
this ratio lies between 0.7 and 1 for most cases, but decreases below 0 for some cases of intense shaking,
which means that instead of sinking, the intruder rises above the grains.
These particular results are a demonstration of the Brazil nuts effect where 
the intense shaking and the friction between grains result in a vertical force
opposed to the weight of the intruder.
This graph demonstartes that the isostatic position is a relatively good approximation for the final position
during H.L.,
although the results are somewhat dispersed, especially at relatively
low acceleration.

%%%%%%%%%%%%%%%%%%%%%%%%%%%%%%%%%%%%%%%%%%%%%%%%%%%%%%%%%%%%%%%%%%%%%%%%%%%%%%%%%%%
\subsection{Penetration dynamics in liquefied cases}
%%%%%%%%%%%%%%%%%%%%%%%%%%%%%%%%%%%%%%%%%%%%%%%%%%%%%%%%%%%%%%%%%%%%%%%%%%%%%%%%%%%

\subsubsection{Data collapse and master curve}
\begin{figure}[htbp]
\begin{center}
        \includegraphics[width=8cm]{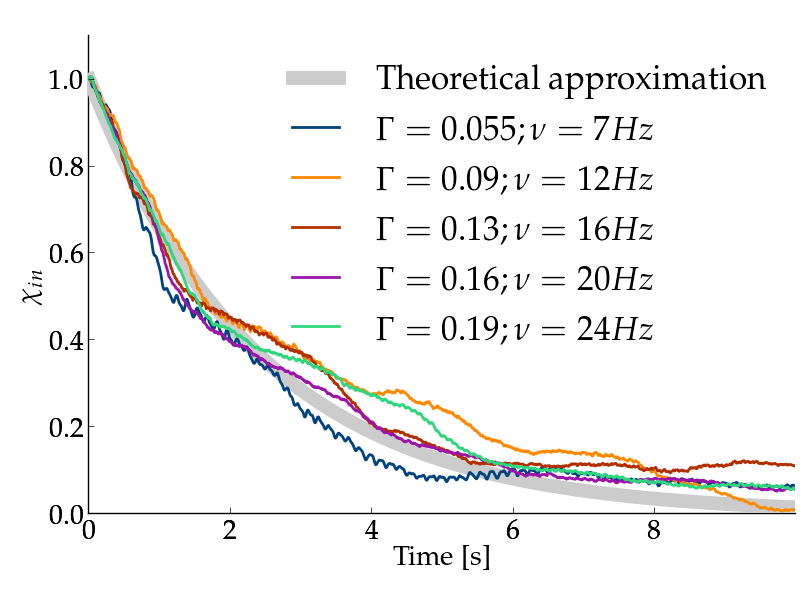}
\end{center}
	\caption[Simulations at different amplitude and frequency with the same velocity peak]
        {Normalized emerged volume $\chi_{in}(t)$ as function of time in simulations at different amplitudes and frequencies, but with an approximatively constant PGV  of 0.012 m/s.
        It can be seen that simualtions ran with the same PGV follow the same master curve.
	The theoretical relaxation of the intruder for a PGV of 0.012
        m/s, calculated in part \ref{Theoretical point of view}, is plotted in a thick gray curve.
        }
        \label{fig:master_meme_pgv}
\end{figure}

\begin{figure}[htbp]
\begin{center}
	\includegraphics[width=8cm]{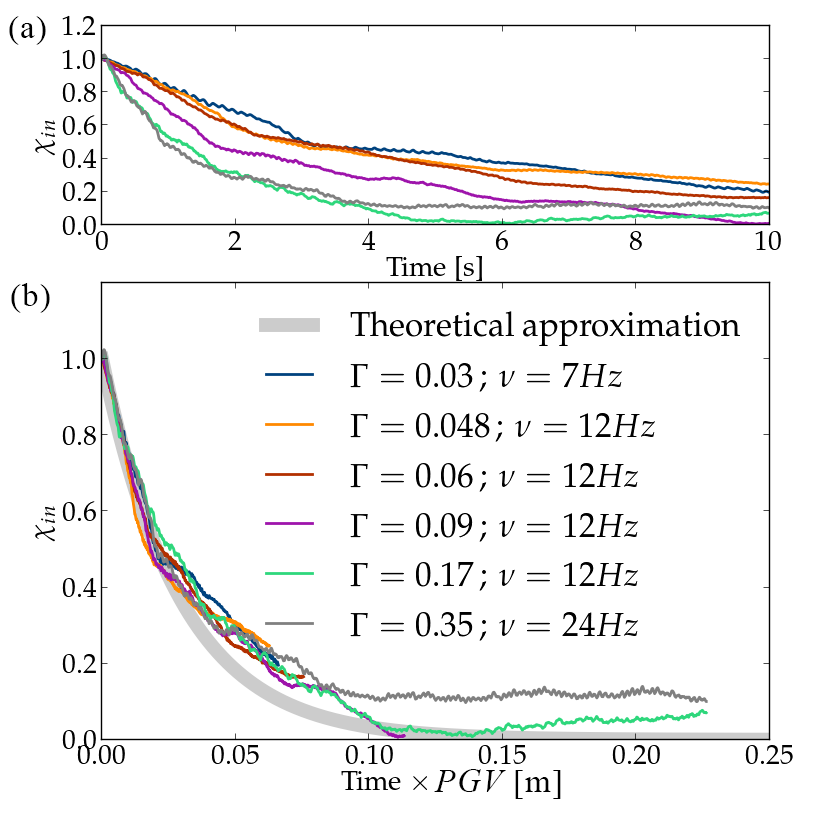}
        \caption[]
	{(a) Normalized emerged volume of the intruder 
	as function of time $\chi_{in}(t)$ in a set of simulations; (b) Normalized emerged volume as function of a normalized time, multiplied
	by the PGV of each respective run, for the same set of simulations. The curves show a reasonable collapse on a master curve.
	In thick grey we add the theoretical relaxation of the intruder.
	The calculation is made in part \ref{Theoretical point of view}.
}
      \label{fig:alignement_final}
\end{center}
\end{figure}

%%%%%%%%%%%%%%%%%%		EXEMPLES	%%%%%%%%%%%%%%%%%%
To understand further the phenomenon of sinking of an object inside a liquefied granular medium,
we investigate the dynamics of the intruder, and how it penetrates towards its equilibrium position in the liquefactions cases.
For simulations made at different amplitudes and frequencies
but at the same peak ground velocity (PGV),
one can observe that all the curves align together,
see Fig.~\ref{fig:master_meme_pgv}.
This observation guides us to collapse the immersed volume vs time curves,
and establish a master curve followed by all the simulations.
Considering simulations made with any amplitude and frequency, whose sinking vs time curves are shown on Fig.~\ref{fig:alignement_final} on top,
we are able to collapse all curves of evolution of the sinking depth
by plotting it as function of a reduced time corresponding to the cumulated strain  imposed by the oscillations, i.e. the time multiplied by the PGV: see Fig.~\ref{fig:alignement_final}(b).
This shows that the speed of penetration of the intruder mainly depends on the peak velocity of the shaking.

%%%%%%%%%%%%%%%%%%		EXPONENTIAL SHAPE	%%%%%%%%%%%%%%%%%%
\subsubsection{Exponential relaxation and characteristic time}
%%%%%%%%%%%%%%%%%%		EXPONENTIAL SHAPE	%%%%%%%%%%%%%%%%%%

Concerning the shape of the sinking curves, a naturally expected shape is an exponentially or a logarithmic decreasing function.
Indeed, on one hand linear systems relax towards equilibrium following an exponential evolution.
On the other hand, in non linear systems close to jamming or pinning,
slow relaxation or creep dynamics often lead during long time to a deformation logarithmic in time.
This is for example the case for dry grain packings compacting under vibrations, 
\cite{Knight1995,Richard2005}, for creep in fracture  propagation \cite{Lengline2011} or for deforming rocks \cite{Brantut2013}.
We use semi-logarithmic representations to verify if the sinking results reveal one of these behaviors.
\begin{figure}[htbp]
	\begin{center}
%	A)Logarithmic fit attempt\\
	\includegraphics[width=7cm]{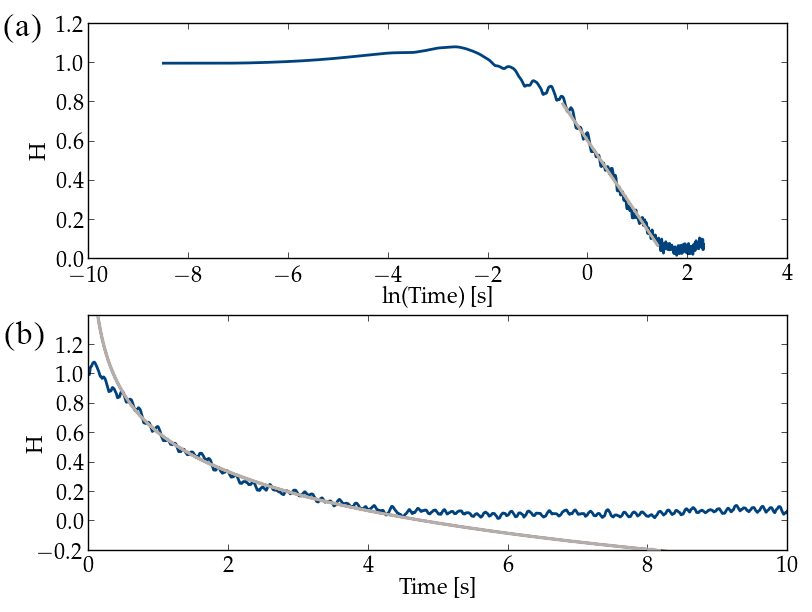}
\hfill
%	B)Exponential fit attempt\\
	\includegraphics[width=7cm]{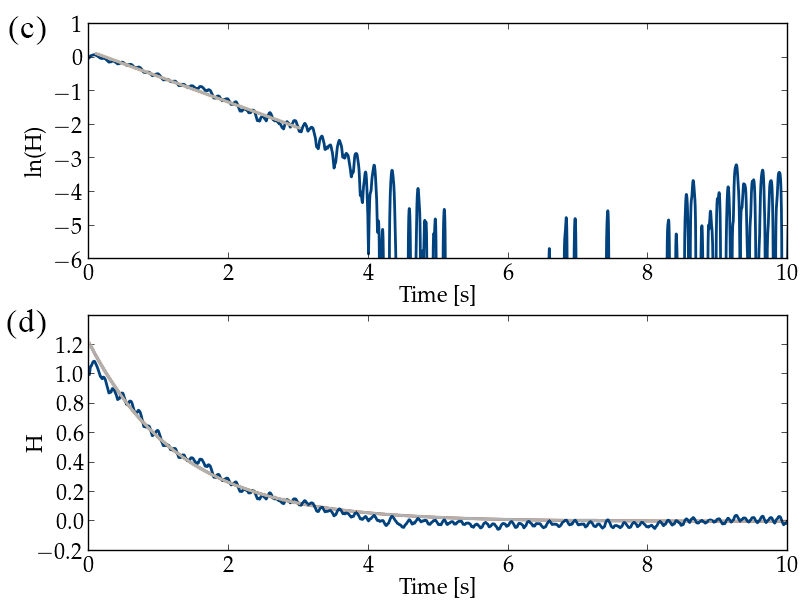}
\caption[Decreasing laws of the emerged height of the intruder]
	{Descent laws for intruders: fitting the normalized emerged height of the intruder $H(t)$ as function of time.
	We apply a logarithmic fit to the simulation data (dark blue curve)
        in (a), (b) and an exponential fit in (c), (d).	Fits are shown
        in light gray. Sinking of the intruder is
        best modelled by an exponential decrease of the height.
	}
\label{fig:expfitex}
\end{center}
\end{figure}
Fig.~\ref{fig:expfitex} displays the sinking of our intruder during one simulation -- characteristic of the majority of the simulation cases.
Fig.~\ref{fig:expfitex} (a)  shows an attempt to a logarithm fit to the 
sinking of the intruder,
and  Fig.~\ref{fig:expfitex} (b) shown an exponentially decreasing fit attempt to the same data. 
For both cases we presented the result first with a semilogarithmic scale,
where an exponential behavior would correspond to  a straight line, and then with a linear scale.
The exponentially decreasing function fits the vast majority of the liquefied simulations
whereas the logarithmically decreasing ones fit only few simulations whose behavior
was hard to distinguish between rigid and liquefied - these correspond to the few slow liquefaction cases that will be developed in further detail in Sec.~\ref{slowliq}.
%IL FAUDRAIT DEVELOPPER, ET MONTRER QUE LES EXPS SONT VALABLES LA PLUPART DU TEMPS, A PART POUR LA SLOW LIQUEF.
When the granular medium is well liquefied we can then assume that the intruder follows an exponential 
sinking toward its equilibrium position.
This will be confirmed by a physical explanation later on.
We apply the exponential fit to every simulation
categorized as liquefied and systematically compute the half-life times.
The procedure is semi-automatic.
We compute the intruder normalized emerged height $H$ defined  as follows:
$H(t) = \frac {h(\infty)- h(t))}{h(\infty)- h(0)}$ with $h(t)$ the immersed height previously introduced
and $h(\infty)$ the immersed height at isostatic position.
$H$ is defined by the same principle as $\vnorm$  (equation \ref{eq:nev}), i.e. as a normalized height
which starts at 1 for every case and goes to 0 if the final immersed height of the intruder reaches $h(\infty)$, the isostatic position.
We plot $H$ with a logarithmic y-axis, as on Fig ~\ref{fig:expfitex}(c), and 
we pick manually the duration corresponding to a straight line (orange line on the plot).
This duration should not be smaller than 3 seconds.
We then apply a linear regression to the normalized emerged height $H$, using the selected duration
and a logarithmic y-axis, for different values of $h(\infty)$:
the linear regression is applied to 15 values of $h(\infty)$ equally spaced between the isostatic position minus 15\%
and the isostatic position plus 15\%.
We keep the result which gives the largest correlation coefficient.
Eventually, from the slope of the linear regression $\lambda$, we obtain the half life time as $t_{1/2}=\ln(2)/\lambda$.
%% RENAUD : obtenu de la définition de demi vie, de exp(-lambda t_{1/2}) = ½ : il faut donc multiplier l’inverse de la pente par ln(2))
%\cite{Huerta2005} LES INTRUS DEVRAIT COULER PLUS LOIN SI C'ETAIT DU FLUIDE.

\begin{figure}[hbtp]
\begin{center}
        \includegraphics[width=8cm]{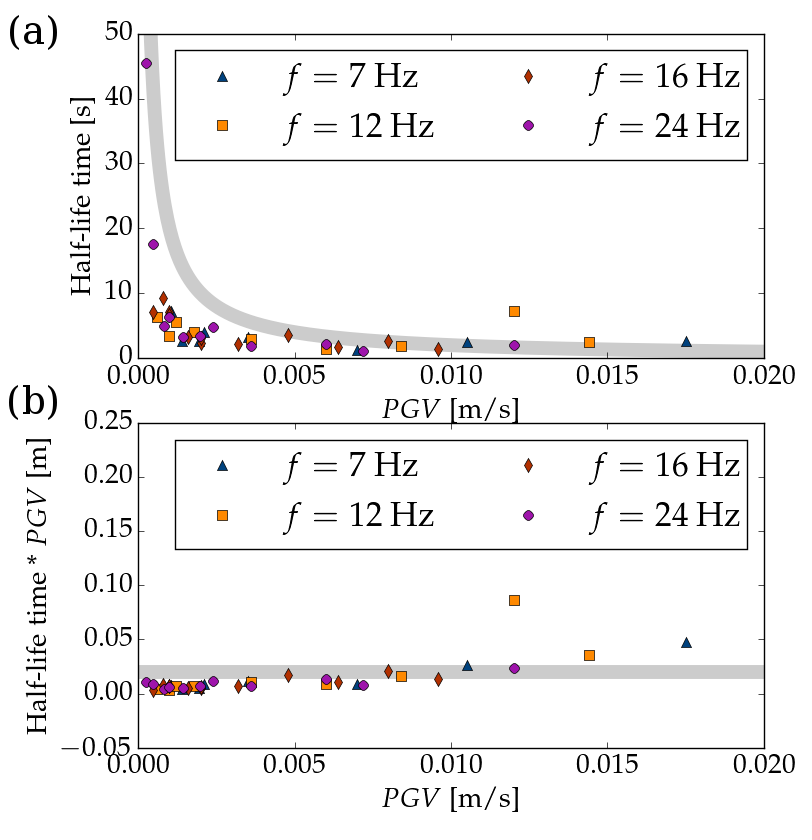}
%        \label{fig:expfit}
\end{center}
        \caption[Half-life time as a function of the PGV]
        {(a) Half-life time for sinking of intruders in simulations, as a function of the peak ground velocity (PGV).
                The shapes of the markers correspond to the shaking frequency.
	The thick grey line plots theoretical values for the half-life time according to the calculation
	 made in part \ref{Theoretical point of view}, i.e. $t_{1/2}=0.02 \mathrm{m}/$PGV.
	(b) $PGV \times t_{1/2}$, product of the Half-life time by the
        PGV, as a function of the PGV. Part \ref{Theoretical point of
          view} predicts that this product is a constant. 
        }
	\label{im:tau}
\end{figure}
We are now able to plot the half-life time of the simulations according to the PGV of the imposed shaking,
see Fig.~\ref{im:tau}.
Different markers are used for  different frequencies.
It is clear from the collapse of Fig.~\ref{fig:alignement_final} and from the 
half-life time dependency of Fig.~\ref{im:tau} 
%(PEUT ETRE AVANT, UNE FIG SANS COLLAPSE?)
that the half-life time is proportional to the inverse of the PGV as 
all the points follow the same master curve regardless of the frequency.
At first order the master curve is the inverse function.
We can now present a physical explanation for the exponential sinking and for the
half-life times dependence on the PGV.
%(DONNE LE FIT OBSERVE POUR depth en fonction de t, i.e. la courbe maitresse avec le temps de vie en fonction des parametres)

%%%%%%%%%%%%%%%%%%%%%%%%%%%%%%%%%%%%%%%%%%%%%%%%%%%%%%%%%%%%%%%%%%%%%%%%%%%%%%%%%%%
\subsection{Theoretical point of view}
\label{Theoretical point of view}
%%%%%%%%%%%%%%%%%%%%%%%%%%%%%%%%%%%%%%%%%%%%%%%%%%%%%%%%%%%%%%%%%%%%%%%%%%%%%%%%%%%

The exponential descent of intruders into granular media has already been reported in a related study by \cite{Sawicki2009}.
They used real sand and a steel cylinder as an intruder.
To understand the origin of this exponential behavior,
we will make an approximate mechanical analysis of the granular medium rheology.
There are mainly two forces acting on the intruder, apart from its weight:
A buoyancy force coming from the fluid and a frictional force
exerted by the solid contacts with the granular medium.
A recent study \cite{BrzinskiIII2013a} shows that in dry granular media
shaken horizontally, the frictional force opposing
the intruders sinking, acts locally normal to the intruder surface.
Following the authors of that study, we assume that this stays valid in the saturated case, and since the penetration speed is low, we assume that the frictional force is proportional
to the speed of the intruder in the medium.
%REF NEEDED.
%%%%%%%%%%%%%%%	CALCUL DEBUT	%%%%%%%%%%%%%%%%%%%%%%%%%%%%%%%%%
%%%%%%%%%%%%% somme des forces = ma %%%%%%%%%%%%%%%%%%%%%%%%%%%%%
Newton's second law applied to the intruder and projected on the vertical axis
can be phrased as:
\begin{equation}
\din V \frac{d^2\h}{dt^2} = -\din Vg + \deff V_{\text{im}}(\h)g -\alpha \frac{d\h}{dt},
\label{equadiff}
\end{equation}
where $u$ is the vertical displacement of the intruder, $\din$ its density $V$ its volume,
and $\deff$ the effective density of the saturated granular medium.
We checked that $\rho_B V \frac{d^2\h}{dt^2}$ is very small relative to the other forces.
In this case, the equation of motion can be simplified as:
\begin{equation}
\frac{d\h}{dt} = \frac{1}{\alpha} ( -\din Vg + \deff V_{\text{im}}(\h)g ).
\label{equadiff_approx}
\end{equation}
One solution of equilibrium exists when $\din<\deff$.
When this solution, called $\heq$, is reached, the term $\frac{d\h}{dt}$ is equal to 0.
Thus
$$
\deff V_{\text{im}}(\heq) = \din V.
$$
%%%%%%%%%%%%  approximation pres de letat dequilibre %%%%%%%%%%%%%%
We focus on the dynamics near the equilibrium state,
in which case:
$$
V_{\text{im}}(\h) = V_{\text{im}}(\heq) + \frac{dV_{\text{im}}}{d\h}|_{\heq} (\h-\heq)
$$
The term $\frac{dV_{\text{im}}}{d\h}|_{\heq}$ is linked to a characteristic surface of the intruder.
%It represents the cylinder of infinitesimal height $d\h$, in the intruder around the surface of the medium.
%Its basis is the disc of intersection between the intruder at equilibrium position and
%and the effective fluid surface.
It represents the disc of intersection between the intruder at equilibrium position and the effective fluid surface.
As the equilibrium position is close to total immersion,
the intersection between the intruder and the effective fluid surface decreases when the
intruder approaches its equilibrium,
and so $\frac{dV_{\text{im}}}{d\h}|_{\heq}$ is a negative term.
Thus we will write $\beta =-\frac{dV_{\text{im}}}{d\h}|_{\heq}$, where $\beta$ is positive.
Eq.~(\ref{equadiff_approx}) becomes:
$$
\frac{d\h}{dt} = \frac{1}{\alpha} \left( -\din Vg + \deff (V_{\text{im}}(\heq) - \beta (\h-\heq))g \right).
$$
Thanks to the expression of the equilibrium solution $V_{\text{im}}(\heq)$, this becomes:
$$
\frac{d\h}{dt} = \frac{\beta g \deff}{\alpha} (\heq-\h).
$$
We finally reach a linear differential equation.
Using $\hini = \h(t=0)$, the initial position, we obtain the following solution for $\h$:
\begin{equation}
\h(t) = (\hini-\heq)e^{\frac{-\beta g \deff}{\alpha}t} + \heq.
\label{eq_decroissance_exp}
\end{equation}
So finally, after having assumed a negligible acceleration for the intruder and a frictional force
proportional to the velocity of the intruder,
we find that the movement of the intruder around its equilibrium position
follows an exponentially decreasing law.

Concerning the half-life time of penetration, our results allow to provide an expression for $\alpha$.
In Eq.~(\ref{eq_decroissance_exp}), we have an expression for the half-life time $t_{1/2}$:
\begin{equation}
t_{1/2} = \frac{\ln(2)\alpha}{\beta g \deff}.
\label{eq:t12lettre}
\end{equation}
We can measure a particular half-life time with figure \ref{fig:master_meme_pgv}.
The simulations plotted in this figure have a PGV of $V_{0}=0.012$ m/s and the half-life time is equal to $T_{0}=1.7$ s.
We can assume that 
\begin{equation}
\label{eq:tdemi1}
t_{1/2} = T_{0}\,F(A\, 2\pi f)
\end{equation}
where $F$ is a function of  $(\, 2\pi Af)$ the PGV of the simulations
with $F(V_{0})= 1$.
Then we use the observation made in figure \ref{fig:alignement_final} and \ref{im:tau},
namely that for simulations having different PGV, 
the alignement is obtained by multiplying the time axis by the PGV.
The interpretation is that the caracteristic time of the decreasing of $\vnorm$,
the half-life time in the case of an exponential decrease,
is a function of the form 
\begin{equation}
\label{eq:tdemi2}
t_{1/2} = \frac{C}{A\, 2\pi f}
\end{equation} 
where $C$ is a constant.
We can equate the two expressions \ref{eq:tdemi1} and \ref{eq:tdemi2} in order to find $C=T_{0}V_{0}=0.02 \mathrm{\ m}$.
In figures \ref{fig:alignement_final} and \ref{fig:master_meme_pgv},
we have drawn the theoretical lines corresponding to an exponential decrease from 1 to 0
with the caracteristic time defined by the half-time $t_{1/2}$ of this last Eq.~(\ref{eq:tdemi2}).
The agreement with the halftimes extracted from the simulations
demonstrates the consistency of this expression.
Combining expressions \ref{eq:t12lettre} and \ref{eq:tdemi2} provides
an expression for $\alpha$, which determines the prefactor of the friction law for the penetration of the intruder in the shaken granular medium, as function of $\beta$, the cross section of the intruder close to the isostatic position, the density of the medium and the PGV $A 2 \pi f$:
$$\alpha = \frac{\beta g \deff}{\ln(2)}   \frac{T_{0}V_{0}}{A\, 2\pi f}$$
%A COMPARER AUX MASTER CURVES, ET A T1/2 EN FONCTION DES PARAMS?
$\alpha$ may be viewed as an effective viscosity coefficient.

%%%%%%%%%%%%%%%%%%%%%%%%%%%%%%%%%%%%%%%%%%%%%%%%%%%%%%%%%%%%%%%%%%%%%%%%%%%%%%%%%%%
\subsection{Particular case of slow liquefaction following logarithm penetration}
\label{slowliq}
%%%%%%%%%%%%%%%%%%%%%%%%%%%%%%%%%%%%%%%%%%%%%%%%%%%%%%%%%%%%%%%%%%%%%%%%%%%%%%%%%%%

\begin{figure}[htbp]
\begin{center}
        \includegraphics[width=6cm]{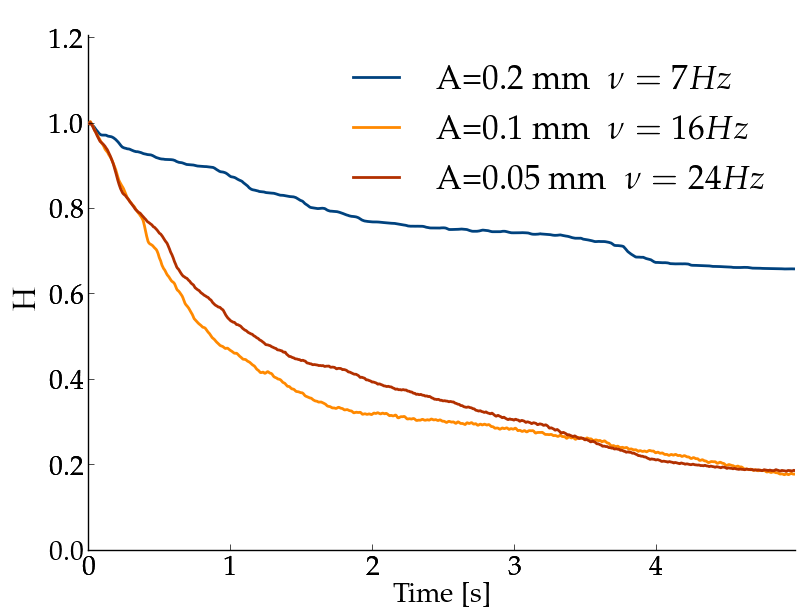}
	\caption{Normalized height as function of time in simulations with identical peak ground velocity.
	The simulation made at the frequency 7Hz does not align with the two other.
	It is a case of slow liquefaction.}
        \label{meme_pgv_lique_lente}
\end{center}
\end{figure}

The dynamics of the intruder described above predicts the behavior of most of the
liquefied (H.L.) simulations and experiments, yet
%DIT DABORD QUE  CA MARCHE DANS LA PLUPARt DES CAS, MAIS... fails to explain the
fails to reproduce the behavior in a few cases.
Indeed among the simulations classified as liquefaction state,
there are some few cases where the sinking
of the intruder does not follow the same trend as described above.
When the intruder movement is compared to other simulations ran at the same PGV,
the curve does not align with the other ones although they all
have a liquefaction behavior, see Fig.\ref{meme_pgv_lique_lente}.

\begin{figure}[htbp]
\begin{center}
%        A)Logarithmic fit attempt\\
        \includegraphics[width=8cm]{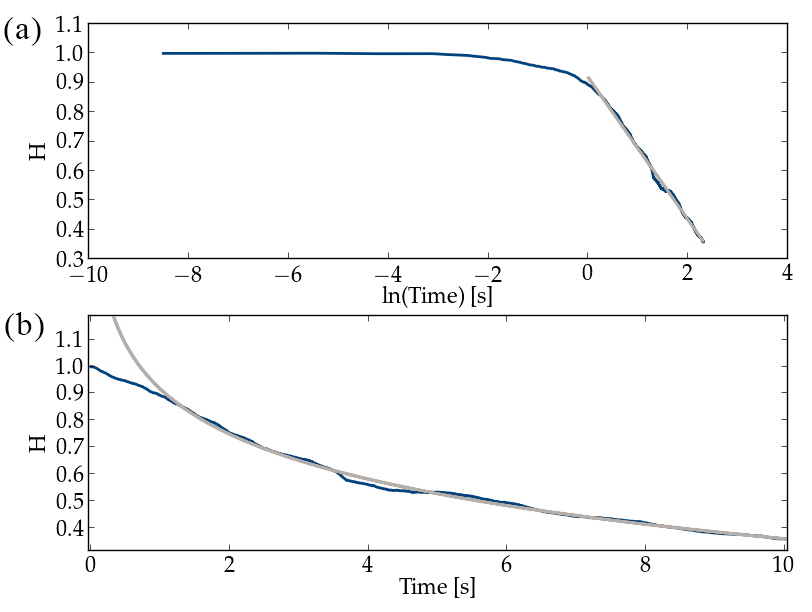}
%        B)Exponential fit attempt\\
        \includegraphics[width=8cm]{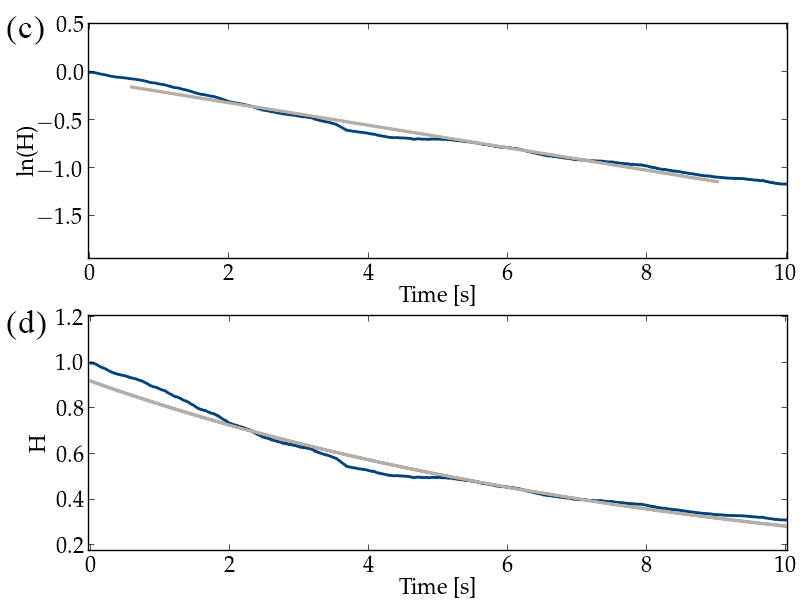}
        \caption[Decreasing laws of the emerged height of the intruder]
        {Descent laws for intruders: fitting the normalized emerged height of the intruder $H(t)$ as function of time.
        We apply a logarithmic fit of the data in (a), (b) and an exponential fit in (c), (d). Fits are in light gray, data is in dark blue.
        We find that in some (less common) cases the emerged height follows a logarithmic decrease.
        }
        \label{im:logfit}
%       \caption[Exponential decrease of the emerged volume of the intruder]
%       {Exponential decreasing of the emerged volume of the intruder.
%               On the top we choose a logarithmic vertical scale.
%               by interpolating the data we obtain a strait line.
%               On the bottom the vertical scale is linear and
%               we observe the emerged volume and its modeled exponential decreasing.
%       }
\end{center}
\end{figure}

We call these particular cases slow liquefaction.
Observing the dynamics of penetration of the intruders in these slow liquefaction cases, we
find that a logarithmic law fits better than the exponential one,
as shown on Fig.\ref{im:logfit}.
This is in contrast with most cases of liquefaction where the exponential decrease fits better.
With the numerical simulations we found out that when the size of the numerical box is increased,
these cases get less frequent.
With 2000 particles, the area in the phase diagram where slow liquefaction
appears is very narrow.
%IL FAUDRAIT DRE EN GROS QUAND ON L'OBSERVE.
This behavior agrees with recent studies on granular compaction \cite{Richard2005},
where under small excitation, granular systems get "jammed" and "aging" phenomena are observed.
The stationarity of these systems is typically not reached and logarithmic relaxation is found.
%phrase copiée dans:
%http://www.nature.com/nmat/journal/v4/n2/pdf/nmat1300.pdf
Our explanation for this logarithmic dynamic is that the system needs to explore rearrangement of larger and larger 
amount of particles.
This can be compared to the glassy dynamic behavior and the parking lot model \cite{Kolan1999}.
%http://journals.aps.org/pre/abstract/10.1103/PhysRevE.59.3094

%%%%%%%%%%%%%%%%%%%%%%%%%%%%%%%%%%%%%%%%%%%%%%%%%%%%%%%%%%%%%%%%%%%%%%%%%%%%%%%%%%%
\section{Discussion and Conclusions}
%%%%%%%%%%%%%%%%%%%%%%%%%%%%%%%%%%%%%%%%%%%%%%%%%%%%%%%%%%%%%%%%%%%%%%%%%%%%%%%%%%%
%DIT DABORD QUEL EST LE MODELE THEORIQUE ET SES PREDICTIONS
Using a model of granular soil and a sphere representing a structure built on the top of this soil,
we are able to reproduce soil liquefaction by shaking the medium.
Indeed, with sufficiently strong shaking, the sphere, originally positioned on top of the medium, sinks quickly into it.
Using basic physics equations taking into account buoyancy and friction between the grains,
we construct a theoretical model which predicts under which imposed shaking accelerations the sphere will sink and what will be
its final position.
This theoretical model predicts three different regimes for the shaken medium,
depending on the friction coefficient of the material, the presence of water and the density of the grains.
When increasing the imposed acceleration, we expect subsequently, a rigid regime,
a heterogeneous liquefaction regime allowing the sphere to sink
(representative of the conditions during seismically triggered liquefaction in nature),
and a global excitation liquefaction regime where the whole medium strongly deforms. 
Our theoretical model to explain liquefaction has been validated by numerical simulations and experiments.
Our experiments consist of 3D granular media, composed of light monodisperse beads,
horizontally shaken by regular oscillations.
Our simulations implement a molecular dynamics code in 2D with gravity, buoyancy and contact forces.
In our systems, liquefaction is controlled by the competition between the buoyancy and the gravitational forces.
Because the buoyancy isn't applied everywhere but only on the immersed grains,
it enables the grains of the basic medium to slide on each other, while the intruder doesn't slide on the surrounding grains, which is at the origin of the heterogeneous liquefaction behavior we observe.
We explored a broad range of accelerations and frequencies.
The main conclusion is that only the peak ground acceleration of the shaking
and material parameters (friction, density, saturation) 
determine the behavior of the saturated granular medium.
%RENAUD: REF TO TROIS REGIMES, QU'EST CE QUI DETERMINE LES DEUX LIMITES, INFLUENCE DE L'EAU (DISPARITION DE LA BANDE DE LIQUEF AVEC LA PROFONDEUR?)
Next, we show that in the case of heterogeneous liquefaction, 
a spherical intruder lying on top of a granular medium
sinks down to a position close to the one set by isotasy in this medium.
Another conclusion relates to the dynamics of the intruder penetration.
We first show a clear data collapse in time among the sinking of the sphere with simulations,
using a normalization by the inverse of the peak ground velocity.
We then show that the intruder displays usually an exponential relaxation towards equilibrium.
We give a theoretical interpretation of this relaxation, and find a relationship between
the effective viscosity of the fluidized medium and the shape of the immersed
volume of the intruder near to its isostatic position.

In a more general way, 
the experiments we performed show that liquefaction is possible 
under drained conditions. 
This is in agreement with recent studies \cite{Goren2010,Goren2011,Lakeland2014}.
Our simulations and experiments in effect suggest an alternative to the common view
(e.g. \cite{Youd2001,Seed2003}) that fluid pressure in pores between
grains must rise beyond the hydrostatic value
in order to produce soil liquefaction. 

Our model is valid in the case of small compaction and permeable media. 
%Indeed for large system with low compaction and high permeability as clay,
%% einat -this is wrong. i think you mean large compaction and low permeability, which is the case of clay ? 
%
Indeed for systems with large compaction and low permeability, such as clay,
the dynamic pressure in pores will not be negligible any more and
viscous forces need to be added in our model as in \cite{Niebling2012a} and \cite{Niebling2012b}.

All the experiments and simulations that we show are made using light particles of bulk density 1050 $\mathrm{kg\cdot m^-3}$, and in a non cohesive saturated medium,
which increases the effect of liquefaction and allows clear depth-of-sinking measurements. Some adaptations are needed to evaluate the liquefaction potential of a real soil. The range of accelerations for which liquefaction occurs will decrease with the use of characteristic soil particle density, and the isostatic position of an intruder in a denser medium will be more emerged than in our experimental (and numerical) media. The presence of cohesion between the particles and the fact that water often does not saturate soils up to the surface will also increase the critical acceleration needed to liquefy a soil. Moreover the monodisperse spherical particles of our media is a chosen simplification of the setup to improve the reproducibility, but is very simplified with respect to  the diversity of shapes, sizes and densities of the particles composing a natural soil - which will certainly affect the friction parameter, and can give rise to additional complexities. Our work is meant to give an insight to the physical phenomenon by which soil liquefaction is triggered during an earthquake, and many perspectives are open in order to explore widely the extensions of this model - notably, by varying saturation, polydispersity, and variable densities.

As a perspective,
we suggest to study the impact 
of the presence of a second intruder nearby on the intruder's penetration.
Indeed during some liquefaction events,
one can notice that a building can sink or tilt
while the neighboring buildings remain stable.
An assumption is that the weight of one building may stabilize the surrounding soil.
Another interesting line of work
would be to confront our model with a broad range of grain densities, and to systematically vary the level of fluid in the medium, to  check the fluid level impact on liquefaction.
Eventually, it would be interesting to use a finite time of shaking
which corresponds to the typical time of an earthquake.

%Moreover the simplicity of our theoretical model,
%confirmed by experiments and simulations,
%make it easy to use in a large amount of system.
%Indeed, neither the size or the shape of the particles of the granular media are requested for 
%the computation of the threshold accelerations.
%Only the value of the density of the particles, the friction coefficient of the medium and 
%the density of the intruder are mandatory.

%We used a small polydispersity and no polydensity in order to control our experimental media.
%The compaction and the fraction of water were the harder parameter to control.
%With a system containing a lot of polydispersity, we can expect to find 
%the Brazil Nut Effect \cite{Clement2010} \cite{Poeschel1995} \cite{Luding1996}.

%In which case is our model valid?
%In which case the fluid viscosity need to be taken into account?
%Polydispersity effect?
%%%%%%%%%%%%%%%%%%%%%%%%%%%%%%%%%%%%%%%%%%%%%%%%%%%%%%%%%%%%%%%%%%%%%%%%%%%%%%%%%%%
\begin{acknowledgments}
%%%%%%%%%%%%%%%%%%%%%%%%%%%%%%%%%%%%%%%%%%%%%%%%%%%%%%%%%%%%%%%%%%%%%%%%%%%%%%%%%%%
We appreciate helpful discussions with K. J. M\r{a}l\o y, E. Altshuler, A. J. Batista-Leyva,
G. S\'anchez-Colina, V. Vidal, G. Sch{\"a}fer, Amir Sagy, Emily Brodsky and L. Goren. We acknowledge the support of the European Union’s Seventh Framework Programme for research,
technological development and demonstration under grant agreement no 316889 (ITN FlowTrans), of the CNRS INSU ALEAS program, and of the LIA France-Norway D-FFRACT.
We also thank 
Alain Steyer and Laurent Rihouey for outstanding support in building
the setups.
%We thank ... for fruitful discussions.
%We also wish to thank ... the Associate Editor, and an anonymous reviewer for constructive reviews
%that helped greatly in improving the manuscript.
\end{acknowledgments}
 
% \bibliographystyle{spphys}       % APS-like style for physics
% \bibliographystyle{plain}
\bibliographystyle{apsrev4-1}  
%\bibliography{liquefaction-RT}   % name your BibTeX data base
%\bibliography{AdditionnalBibliography}   % name your BibTeX data base


\end{document}